\documentclass[preprint,aps,showpacs]{revtex4}
\usepackage{amsfonts}
\usepackage{amsmath}
\usepackage{amssymb}
\usepackage{graphicx}

\newcommand{\be}{\begin{equation}}
\newcommand{\ee}{\end{equation}}
\newcommand{\bes}{\begin{subequations}}
\newcommand{\ees}{\end{subequations}}
\newcommand{\ben}{\begin{eqnarray}}
\newcommand{\een}{\end{eqnarray}}

\begin{document}

\title{Fermion localization and resonances on two-field thick branes}
\author{C. A. S. Almeida$^{1}$, R. Casana$^{2}$, M. M. Ferreira Jr.$^{2}$,
A. R. Gomes$^{3}$}
\affiliation{$^1$ Departamento de F\'\i sica, Universidade Federal do Cear\'a (UFC), \\
C. P. 6030, 60455-70, Fortaleza, Cear\'a, Brazil\\
$^2$ Departamento de F\'\i sica, Universidade Federal do Maranh\~ao (UFMA) \\
Campus Universit\'ario do Bacanga, 65085-580, S\~ao Lu\'\i s, Maranh\~ao,
Brazil\\
$^3$ Departamento de F\' isica, Instituto Federal de Educa\c c\~ao,
Ci\^encia e Tecnologia do Maranh\~ao (IFMA), 65025-001, S\~ao Lu\'\i s,
Maranh\~ao, Brazil}

\begin{abstract}
We consider $(4,1)$-dimensional branes constructed with two scalar fields $%
\phi$ and $\chi$ coupled to a Dirac spinor field by means of a general
Yukawa coupling. The equation of motion for the coefficients of the chiral
decomposition of the spinor in curved spacetime leads to a
Schr\"odinger-like equation whose solutions allow to obtain the masses of
the fermionic modes. The simplest Yukawa coupling $\bar\Psi\phi\chi\Psi$ is
considered for the Bloch brane model and fermion localization is studied. We
found resonances for both chiralities and related their appearance to branes
with internal structure.
\end{abstract}

 \maketitle


\section{Introduction}

Braneworld scenarios have received great attention from the physical
community after addressing important problems such as gauge hierarchy \cite%
{rs2,add,add2} and the cosmological constant \cite{rub2,ce2}. The
construction of domain walls embedded in a higher dimensional bulk, however,
has a more ancient history in the literature \cite{civ,aka,rub,vis}. As is
well-known, $(4,1)$-D branes can be classified as thin and thick ones. Thin
branes are constructed after introducing a tension term in the action,
localized by a Dirac delta function (see \cite{pm} and references therein).
The mathematical formalism that describes the metric around the brane allows
to obtain several interesting results in the domain of high energy physics
and cosmology. Usually, the issue of field localization in such branes is
addressed with the help of Dirac delta functions, without any clear
subjacent dynamics. On the other hand, thick branes are constructed in a
dynamical way after introducing one or more scalar fields coupled with
gravity \cite{df,ce,g,c}. Thick branes are more natural in the sense that
field and gravity localization can be studied with the introduction of
smooth functions (instead of Dirac ones). Moreover, the thin brane solutions
are recovered in certain limits \cite{bcy,bbg}. One important issue in $(4,1)
$-D braneworld models is to consider spacetimes with a bulk metric that
extends the Randall-Sundrum result \cite{rs1} in a way that gravity can be
localized in the $(3,1)$-D slice \cite{pm,mps,bgl,bh2,bh3}. The spacetime
around the Randall-Sundrum brane is anti-de Sitter ($AdS_{5}$), and several
thick brane models (that present this characteristic asymptotically) have
been constructed.

Braneworld models with one or more scalar fields were constructed in order
to relate gravity localization with brane thickness \cite{bg,bg2}. In
particular, the Bloch brane model \cite{bg2} is composed of two scalar
fields and is the extension for the known Bloch walls \cite{bnrt,b3,bs} in
the context of braneworlds. Bloch branes are stable structures that can be
characterized as branes with interfaces located at the maximum of energy
density. The two interfaces signals the presence of an internal structure in
the brane. The presence of internal structure in topological defects
was firstly considered in Witten's superconducting cosmic strings \cite{wit}%
. There the internal structure is provided by the condensation of a
(charged) field over the defect. In the Bloch brane case, the appearance of
internal structure is concerned with a low value of the coupling constant
that guides the way the two fields interact with each other. For a stronger
interaction, a single peak for the energy density characterizes a brane with
simpler structure. It was also shown \cite{bg2} that in the presence of
gravity the interfaces are located more closely in the extra dimension. As
such branes are able to localize gravity, one interesting question is
investigating how fermionic fields may be localized. In this way, there
arises the expectation that a brane with richer internal structure could
provide new results when compared to a simple kinky brane.

The study of fermion localization on branes \cite%
{hs,rds,m,drt,rpu,bgw,kk,mpt,gmt,lzd,lzz,lzwd,dg} is rich and interesting.
In order to localize fermions in branes one needs a coupling between the
spinors and the scalar fields that form the brane. This is a condition also
present in Jackiw and Rebbi treatment of fermion localization on solitons in
flat space \cite{jr}. In branes, the procedure consists in separating from
the full spinor one scalar coefficient with dependence only on the extra
dimension, leading to a Schr\"odinger-like equation and a probabilistic
interpretation. Depending on the model, one can obtain resonant massive
states.

In particular, the authors of \cite{lzd} analyzed the issue of fermion
localization on a brane constructed with the sine-Gordon potential. That
work considered general fermionic Yukawa couplings between one scalar field
and spinorial fields. It was found that the simplest Yukawa coupling $\bar{%
\Psi}\phi \Psi $ allowed left-handed fermions to possess a zero-mode that
localizes on the brane. The right-handed fermions, on the other hand,
present no zero-mode. The large massive fermionic modes (for both left and
right chirality) are plane waves and thereby are not localized on the brane.

Inspired on the results obtained in \cite{lzd}, we investigate the issue of
fermion localization of fermions on a brane constructed from two scalar
fields coupled with gravity. In Sec. 2, we present the known first-order
equations for a brane model coupled with scalar fields \cite{df}. Sec. 3
deals with the fermionic sector of the model, where we extend the results
from \cite{lzd} for two scalar fields $\phi $ and $\chi $ with general
coupling. Afterwards, we choose the simple coupling $\bar{\Psi}\phi \chi
\Psi $. In Sec. 4, we consider the Bloch brane model and our main results
are presented. The Schr\"odinger-like equation for the model is obtained and
numerically solved. The solutions correspond to massive fermionic modes. We
have found resonances on the brane and have related their appearance with
the simultaneous increasing of the brane internal structure. Our conclusions
are presented in Sec. 5.

\section{Scalar field and metric equations}

Our system is described by the action
\begin{equation}  \label{action}
S=\int d^{4}x\,dy\sqrt{|g|}\Bigl[-\frac{1}{4}R+\frac{1}{2}\partial _{a}\phi
\partial ^{a}\phi +\frac{1}{2}\partial _{a}\chi \partial ^{a}\chi -V(\phi
,\chi )\Bigr],
\end{equation}%
and the metric
\begin{eqnarray}  \label{metric_plane}
ds^{2} &=&g_{ab}dx^{a}dx^{b},  \notag \\
ds^{2} &=&e^{2A}\eta _{\mu \nu }dx^{\mu }dx^{\nu }-dy^{2},
\end{eqnarray}%
where $g=\det (g_{ab})$. Here $a,b=0,1,2,3,4,$ and $e^{2A}$ is the warp
factor. We suppose that $A=A(y),\phi =\phi (y)$ and $\chi =\chi (y)$.
The extra dimension $y$ is infinite, and we have the continuity of the first derivatives of $\phi(y)$%
, $\chi(y)$ and $A(y)$ with respect to $y$ as boundary conditions for
the thick brane. 

The action given by Eq.(\ref{action}) leads to the following coupled
differential equations for the scalar fields $\phi (y)$, $\chi (y)$ and the
function $A(y)$ from the warp factor:
\begin{eqnarray}
\label{eom2_field1} \phi ^{\prime \prime }+4A^{\prime }\phi ^{\prime
} &=&\frac{\partial V(\phi
,\chi )}{\partial \phi }, \\
\chi ^{\prime \prime }+4A^{\prime }\chi ^{\prime } &=&\frac{\partial
V(\phi
,\chi )}{\partial \chi }, \\
A^{\prime \prime } &=&-\frac{2}{3}\,\left( \phi ^{\prime 2}+\chi
^{\prime
2}\right) , \\
\label{eom2_field4}
A^{\prime 2} &=&\frac{1}{6}\left( \phi ^{\prime 2}+\chi ^{\prime 2}\right) -%
\frac{1}{3}V(\phi ,\chi ),
\end{eqnarray}%
where prime stands for derivative with respect to $y$.

With the potential \cite{df},
\begin{equation}
V(\phi ,\chi )=\frac{1}{8}\left[ \left( \frac{\partial W}{\partial \phi }%
\right) ^{2}+\left( \frac{\partial W}{\partial \chi }\right) ^{2}\right] -%
\frac{1}{3}W^{2},  \label{gpot}
\end{equation}%
the first-order differential equations which also solve the equations of
motion are
\begin{eqnarray}  \label{eom1_field1}
\phi ^{\prime } &=&\frac{1}{2}\,\frac{\partial W}{\partial \phi }, \\
\chi ^{\prime } &=&\frac{1}{2}\,\frac{\partial W}{\partial \chi },
\\ \label{eom1_field3}A^{\prime } &=&-\frac{1}{3}\,W.
\end{eqnarray}

\section{Fermionic sector}

Now, we consider a Dirac spinor field coupled with the scalar fields by a
general Yukawa coupling. The action for this sector is
\begin{equation}  \label{S_ferm}
S=\int d^{5}x\sqrt{-g}[\bar{\Psi}\Gamma ^{M}D_{M}\Psi -\eta \bar{\Psi}F(\phi
,\chi )\Psi] .
\end{equation}%
Here we consider the fields $\phi$ and $\chi$ as background fields given as
solutions of the Eqs. (\ref{eom2_field1})-(\ref{eom2_field4}). In other
words, we neglect the backreaction from the spinor field on the brane
solutions.

We change variable from $y$ to $z$ with
\begin{equation}
z=\int_{0}^{y}e^{-A(\xi)}d\xi,  \label{zy}
\end{equation}%
in order to rewrite Eq. (\ref{metric_plane}) and get a conformally flat
metric
\begin{eqnarray}
ds^{2} &=&e^{2A}(\eta _{\mu \nu }dx^{\mu }dx^{\nu }-dz^{2}).
\end{eqnarray}%
This change of variable is usual in problems of gravity localization.
Indeed, for a large class of models, one can achieve a Schr\"odinger-like
form for the equations for metric fluctuations, when decoupled from the
scalar fluctuations in a specific gauge. In the treatment of fermion
localization the same change of variable is used. The equation of motion
corresponding to the action given by Eq. (\ref{S_ferm}) is
\begin{equation}  \label{eom_ferm}
\lbrack \gamma ^{\mu }\partial _{\mu }+\gamma ^{5}(\partial _{z}+2\partial
_{z}A)-\eta e^{A}F(\phi ,\chi )]\Psi =0.
\end{equation}%
We apply the well known procedure of a general chiral decomposition for the
spinor $\Psi$:
\begin{eqnarray}  \label{chiral}
\Psi(x,z)=\sum_n \psi_{Ln}(x)\alpha_{Ln}(z) + \sum_n
\psi_{Rn}(x)\alpha_{Rn}(z),
\end{eqnarray}
where the sum can be over discrete bounded modes or over a continuum of
modes. In this decomposition $\psi_{Ln}(x)$ and $\psi_{Rn}(x)$ are,
respectively, the left-handed and right-handed components of the
4-dimensional spinor field with mass $m_n$. Also, there are two scalars $%
\alpha _{Ln}$ and $\alpha _{Rn}$ that depend only on the extra dimension $z$%
. In order to turn the notation easier, we dropped the index $n$ in which
follows. After applying Eq. (\ref{chiral}) in Eq. (\ref{eom_ferm}), we obtain
two equations for the scalars $\alpha _{L}$ and $\alpha _{R}$:
\begin{eqnarray}
\lbrack \partial _{z}+2\partial _{z}A+\eta e^{A}F(\phi ,\chi )]\alpha
_{L}(z) &=&m\alpha _{R}(z),  \label{alpha1} \\
\lbrack \partial _{z}+2\partial _{z}A-\eta e^{A}F(\phi ,\chi )]\alpha
_{R}(z) &=&-m\alpha _{L}(z).  \label{alpha2}
\end{eqnarray}%
Implementing the change of variables $\alpha _{L}=e^{-2A}L$ and $\alpha
_{R}=e^{-2A}R$ \cite{lzd}, we get Schr\"odinger-like equations for the
wavefunctions $R_m(z)$ and $L_m(z)$ given by
\begin{eqnarray}
\label{HL}
H_{L}L_m(z) &=&m^{2}L_m(z), \\
H_{R}R_m(z) &=&m^{2}R_m(z),
\end{eqnarray}%
with corresponding Hamiltonians 
\begin{eqnarray}
\label{factor_HL}
H_{L} &=&\biggl(-\frac{d}{dz}+e^{A}\eta F\biggr)\biggl(\frac{d}{dz}%
+e^{A}\eta F\biggr), \\
\label{factor_HR}
H_{R} &=&\biggl(\frac{d}{dz}+e^{A}\eta F\biggr)\biggl(-\frac{d}{dz}%
+e^{A}\eta F\biggr).
\end{eqnarray}%
The structure of the Hamiltonians guarantees that $m$ is real, the
absence
of tachyonic modes and the possibility of interpreting $|R_{m}(z)|^{2}$ and $%
|L_{m}(z)|^{2}$ as the probability for finding the right and left massive
modes in a given coordinate $z$. The Schr\"odinger equations can be also
written explicitly as
\begin{eqnarray}
-L_m^{\prime \prime }(z)+V_{L}(z)L_m(z) &=&m^{2}L_m,  \label{VschL} \\
-R_m^{\prime \prime }(z)+V_{R}(z)R_m(z) &=&m^{2}R_m,  \label{VschR}
\end{eqnarray}%
with
\begin{eqnarray}
V_{L}(z) &=&e^{2A}\eta ^{2}F(\phi ,\chi )^{2}-e^{A}\eta \partial _{z}F(\phi
,\chi )-(\partial _{z}A)e^{A}\eta F(\phi ,\chi ),  \label{eqVL} \\
V_{R}(z) &=&e^{2A}\eta ^{2}F(\phi ,\chi )^{2}+e^{A}\eta \partial _{z}F(\phi
,\chi )+(\partial _{z}A)e^{A}\eta F(\phi ,\chi ),  \label{eqVR}
\end{eqnarray}%
named as the Schr\"{o}dinger-like potentials for the fermionic fields.
The Hamiltonians from Eqs. (\ref{factor_HL}) and (\ref{factor_HR})
can be written as $H_L=A^\dagger A$ and $H_R=AA^\dagger$. This shows that $%
H_L$ and $H_R$ are conjugated Hamiltonian of supersymmetric quantum
mechanics and $V_{L}(z)$ and $V_{R}(z)$ are superpartner potentials. In this
way there is a zero mode for $V_{L}(z)$ and the spectra of $V_{L}(z)$
and $V_{R}(z)$ are interrelated \cite{cooper, felice}. 

From now on, we consider the simplest Yukawa coupling $F(\phi ,\chi )=\phi
\chi $ and study the occurrence of massive modes for the known Bloch brane
model with two scalar fields.

\section{The Bloch brane model}

A rich model that leads to branes with internal structure is \cite%
{bnrt,b3,bs}
\begin{equation}
W(\phi ,\chi )=2\phi -\frac{2}{3}\phi ^{3}-2a\phi \chi ^{2},  \label{w}
\end{equation}%
with $a$ being a real parameter ($0<a<0.5$). In this case, the first-order
equations (\ref{eom1_field1})-(\ref{eom1_field3}) can be solved
analytically, leading to the following results \cite{bg2}:
\begin{eqnarray}
\phi (y) &=&\tanh (2ay),  \label{phi} \\
\chi (y) &=&\sqrt{\left( \frac{1}{a}-2\right) }\;\mathrm{sech}(2ay),
\label{chi}
\end{eqnarray}%
and
\begin{equation}
A(y)\!\!=\!\!\frac{1}{9a}\Bigl[(1-3a)\tanh ^{2}(2ay)-2\ln \cosh (2ay)\Bigr].
\label{warp}
\end{equation}

\begin{figure}[tbp]
\includegraphics[{angle=0,width=7cm,height=6cm}]{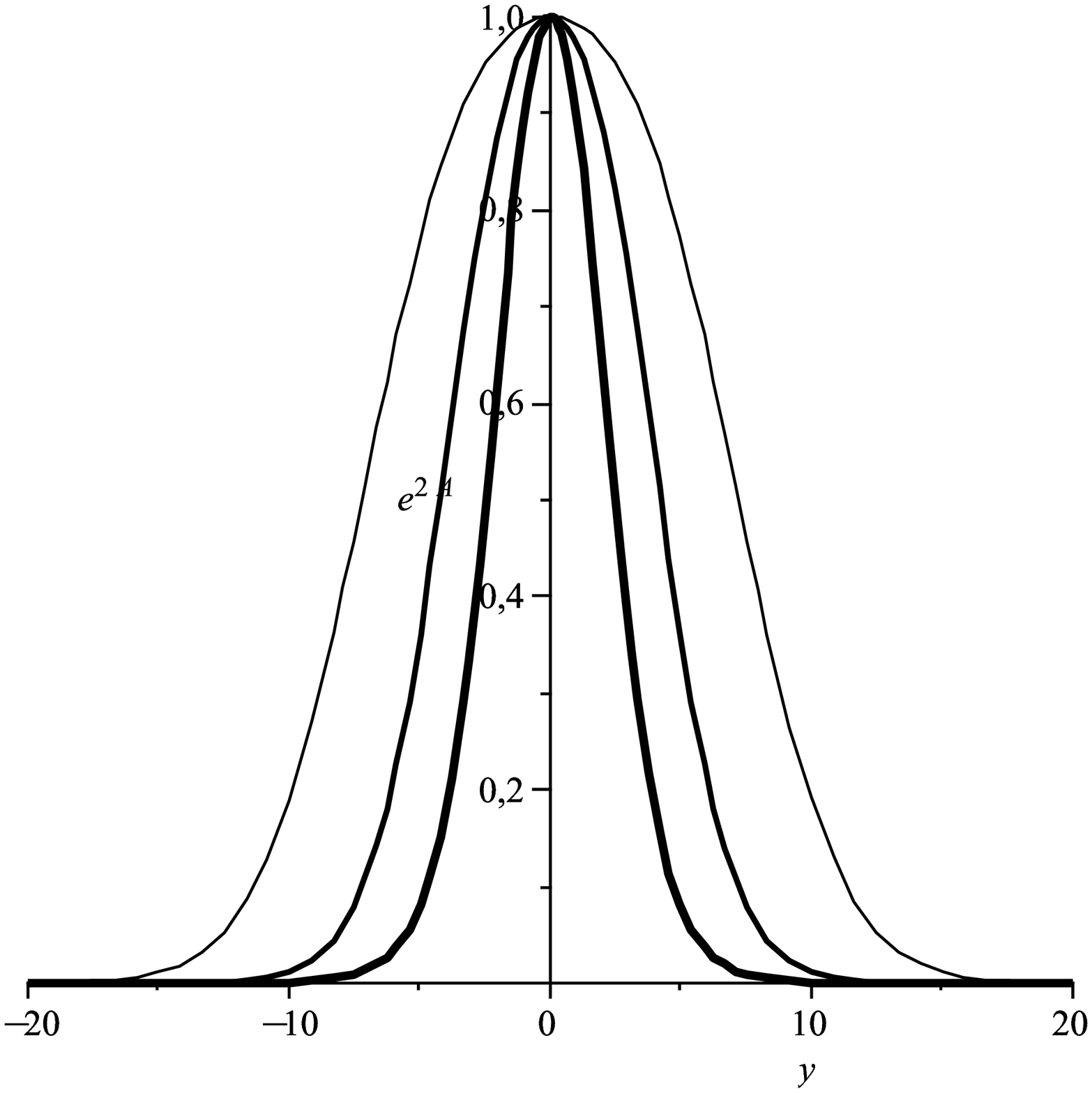} %
\includegraphics[{angle=0,width=7cm,height=6cm}]{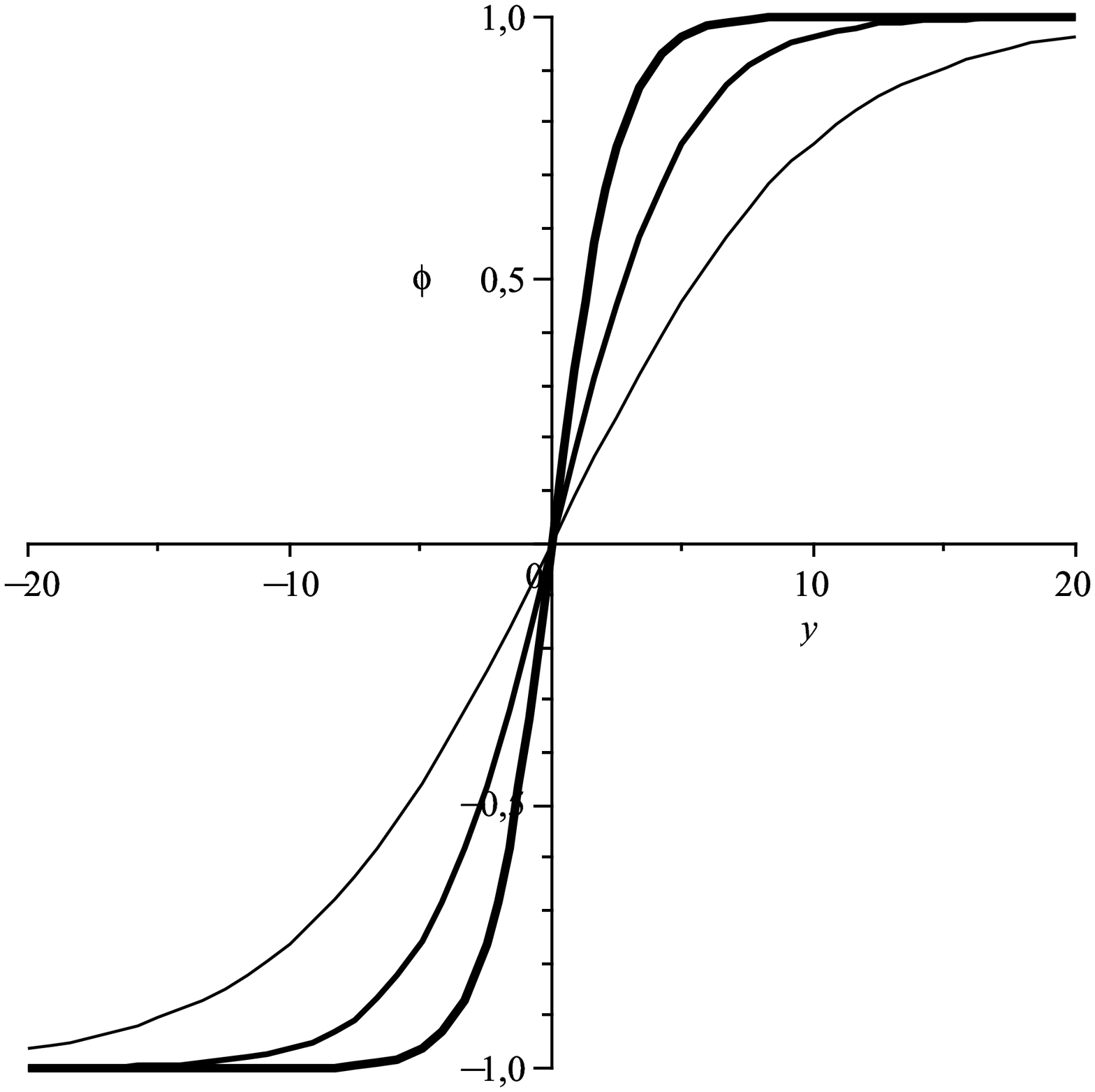}
\caption{(a) Warp factor $e^{2A(y)}$ (left) and (b) field $\protect\phi(y)$
(right) for $a=0.05$ (thinner trace, larger curves), $a=0.10$ and $a=0.20$
(thicker trace, narrower curves).}
\label{figwarp}
\end{figure}

\begin{figure}[tbp]
{\includegraphics[{angle=0,width=7cm,height=6cm}]{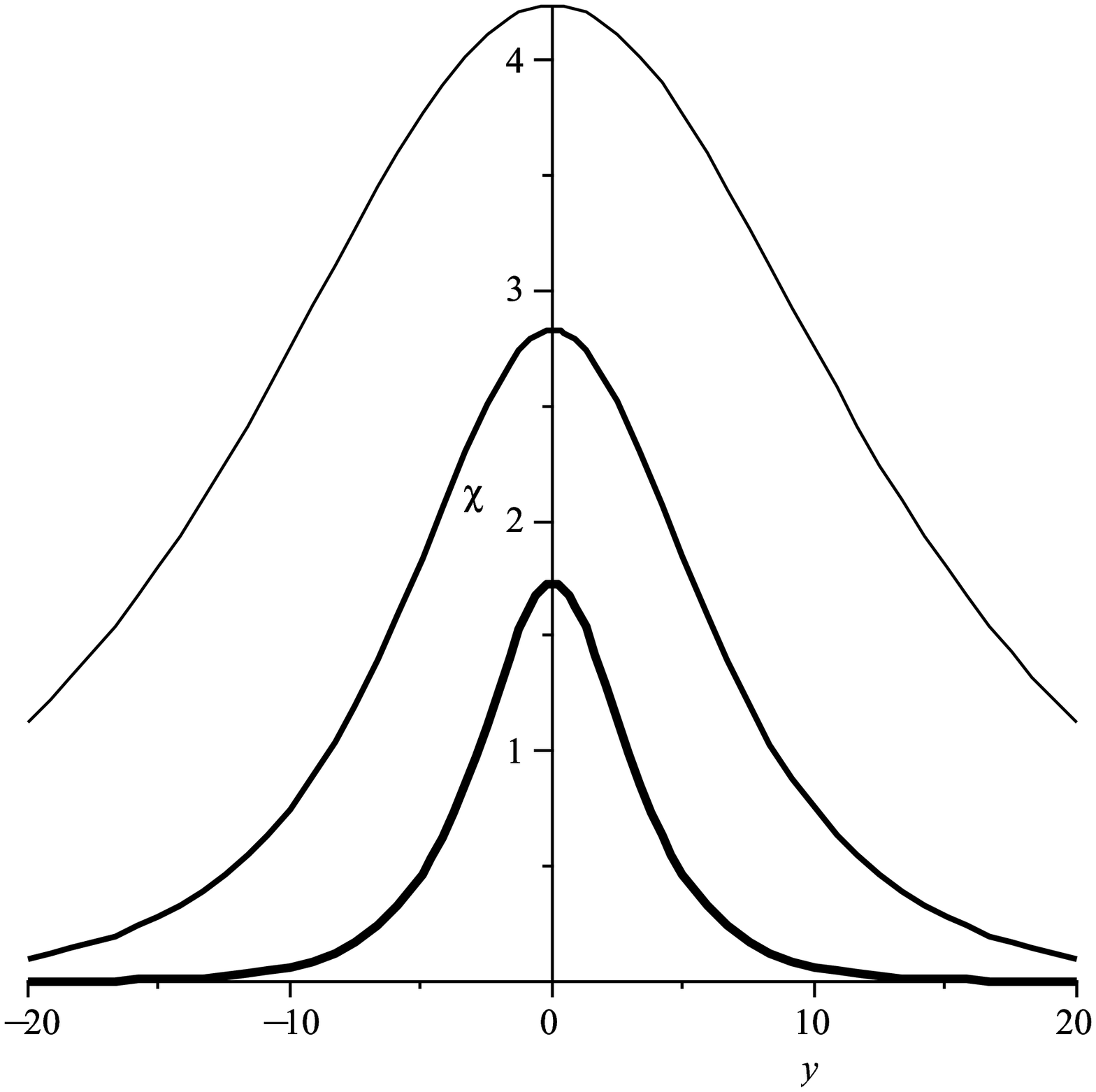} %
\includegraphics[{angle=0,width=7cm,height=6cm}]{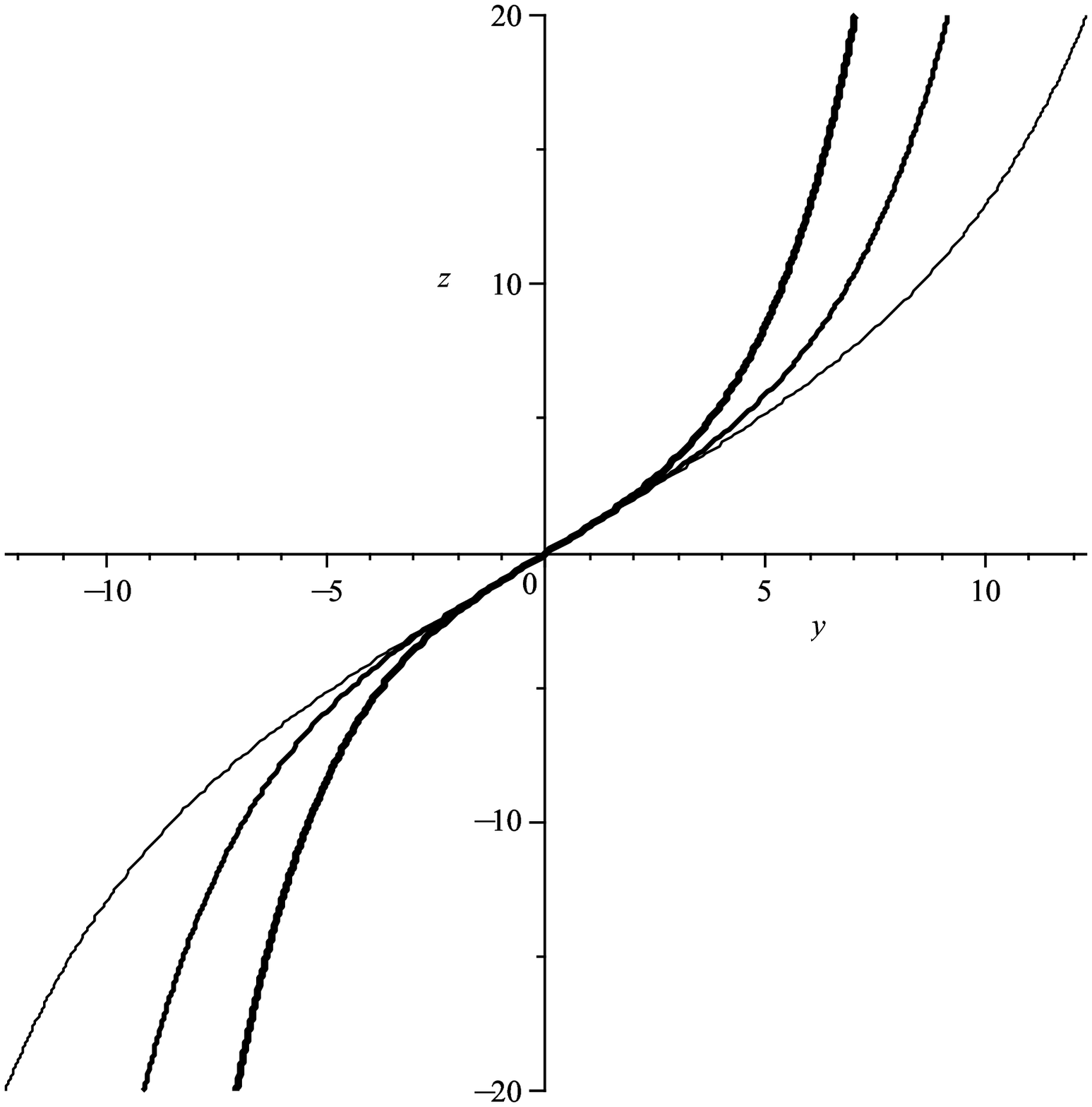}}
\caption{(a) Field $\protect\chi(y)$ (left) and (b) function $z(y)$ (right)
for $a=0.05$ (thinner trace, larger curves), $a=0.10$ and $a=0.20$ (thicker
trace, narrower curves).}
\label{figzy}
\end{figure}

Fig. \ref{figwarp}a shows how the warp factor tends to zero going to
the $AdS_{5}$ limit far from the brane. The brane thickness decreases with $a
$, and this can be seen by the plots of the field solutions in Figs. \ref%
{figwarp}b and \ref{figzy}a.

\begin{figure}[tbp]
{\includegraphics[{angle=0,width=7cm,height=6cm}]{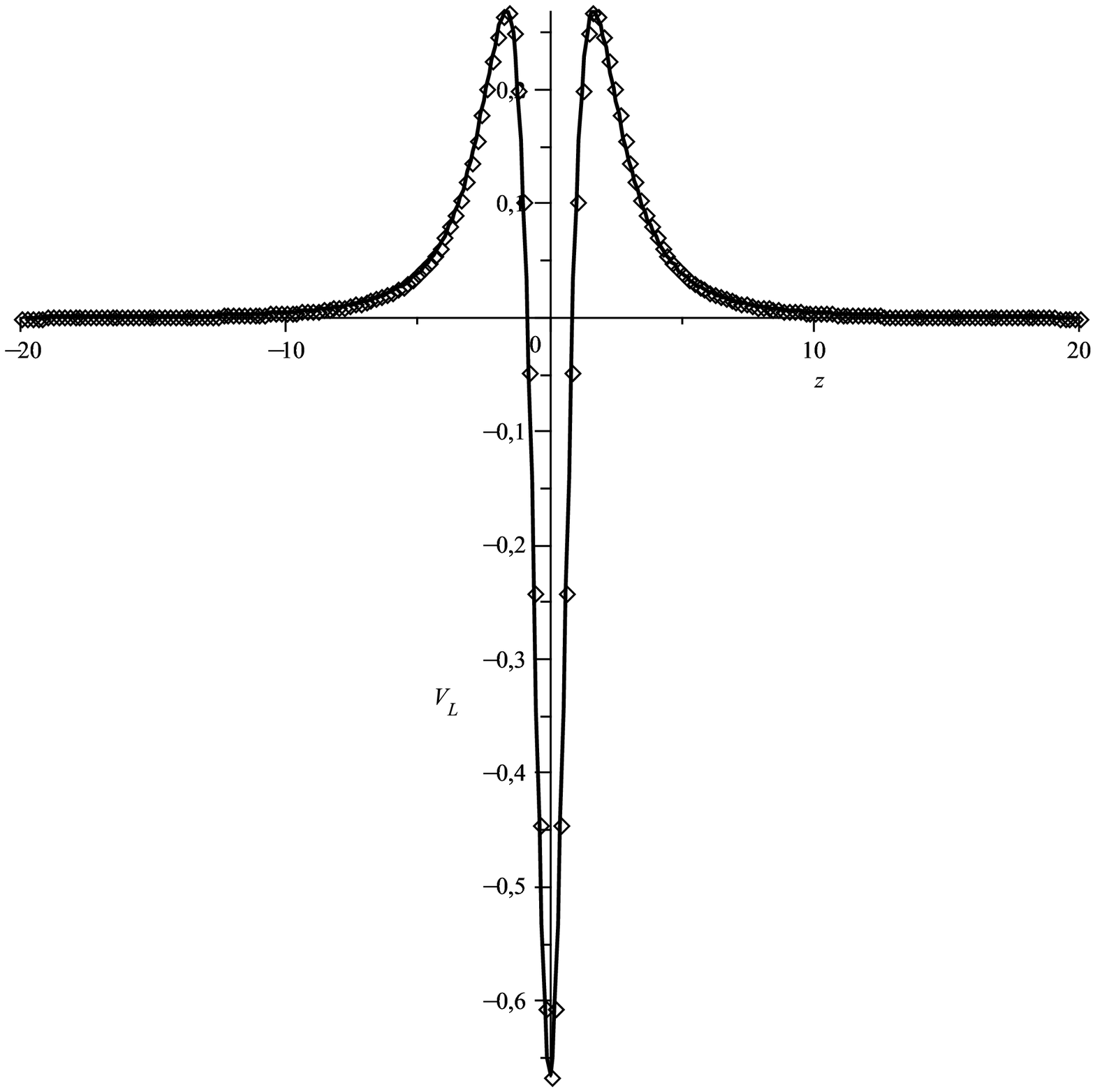}} {%
\includegraphics[{angle=0,width=7cm,height=6cm}]{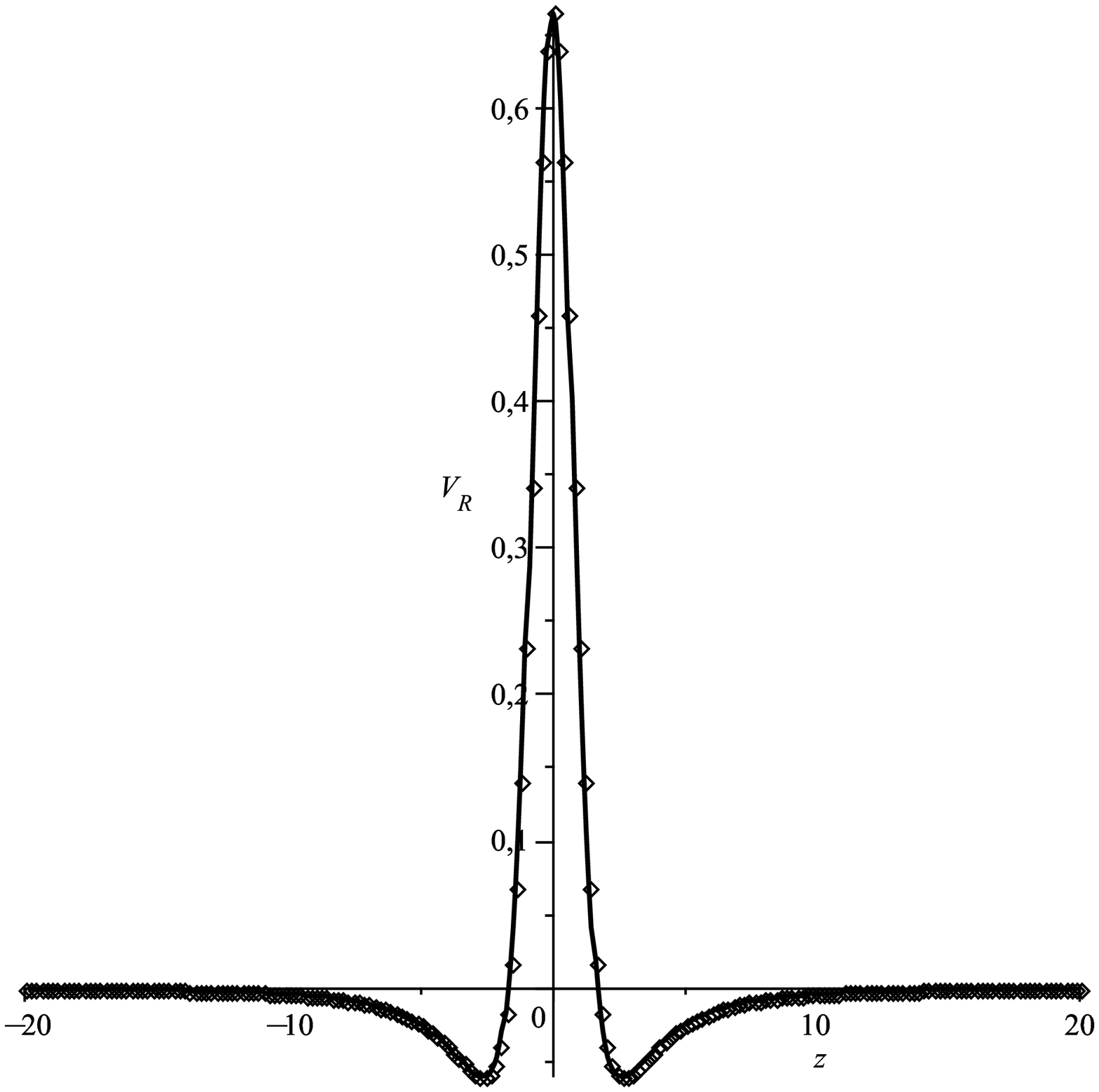}}
\caption{Comparing procedures: Schr\"odinger-like potential (b) $V_L(z)$
(left) and (c) $V_R(z)$ (right) for $a=1/3$, obtained with explicit function
$z(y)$ (points) and with the full numerical procedure (line). Note that both
procedures give almost indistinguishable results.}
\label{figV_explic}
\end{figure}

Now, we turn to the $z$-variable according to Eq. (\ref{zy}). For general
values of $a,$ the expression for $A(y)$ from Eq. (\ref{warp}) is not
suitable to be integrated in a known explicit form and numerical methods are
necessary. However, for the particular case $a=1/3$ we have
\begin{equation}
A(y)\!\!=\!\!-\frac{2}{3}\ln \cosh (2ay),  \label{Ay13}
\end{equation}%
and
\begin{equation}
z=\int_{0}^{y}\biggl[\cosh \biggl(\frac{2}{3}y\biggr)\biggr]^{\frac{2}{3}}.
\end{equation}%
After simple integration, we get an expression for $z(y)$ in an explicit
form in terms of an hypergeometric function:
\begin{equation}
z=\frac{y}{|y|}\left[ \frac{9}{4}\,{{\left( \cosh \left( \frac{2}{3}y\right)
\right) ^{2/3}{}_{2}F_{1}\left( -\frac{1}{3},\frac{1}{2};\frac{2}{3};\left( {%
\cosh }\left( \frac{2}{3}y\right) \right) ^{-2}\right) }}-\frac{9}{8\pi }\,{{%
\Gamma \left( \frac{2}{3}\right) \Gamma \left( \frac{5}{6}\right) }}\right] .
\label{zy13}
\end{equation}%
Unfortunately, even in this simpler case, we can not obtain the inverse $y(z)
$ in an explicit form. However, this explicit form for $z(y)$ is already
useful to produce sequences of pairs of points $(z,y)$ with a constant step
in $z$. Indeed, given a value $z$ the corresponding $y$ can be found after
numerically solving the implicit equation (Eq. (\ref{zy13})). This leads to
the determination of the functions $\phi (z),\chi (z),A(z)$ with good
precision. The corresponding derivative with respect to $z$, $A^{\prime }(z)$%
, can be found as
\begin{equation}
\frac{dA}{dz}=\frac{dA}{dy}\frac{dy}{dz}=\frac{dA}{dy}\left( \cosh {\frac{2}{%
3}y}\right) ^{-2/3},
\end{equation}%
with $dA/dy$ stemming from the explicit expression for $A(y)$ (Eq. (\ref%
{Ay13})). Similarly, we can construct the functions $\phi ^{\prime }(z),\chi
^{\prime }(z)$. All the former calculations can be used to construct the
Schr\"odinger potentials $V_{L}$ and $V_{R}$ displayed in Fig. \ref{figV_explic}
with good precision; this was used with the purpose of validating the
numerical method used for general values of $a$.
.

For some values of $a,$ we integrated numerically Eq. (\ref{zy}) obtaining $%
z(y),$ whose graphic is depicted in Fig. \ref{figzy}b. With this, we can
numerically determine $A(z)$, $\phi (z)$, $\chi (z)$. In order to obtain the
derivatives $A^{\prime }(z),\phi ^{\prime }(z),\chi ^{\prime }(z)$, a
numerical procedure was constructed for generating sequences with constant
step in $z$. In this way the procedure for general values of $a$ is more
involved once we do not have an explicit form for $z(y)$ to guide us as in
the $a=1/3$ previous case. Eqs. (\ref{eqVL}) and (\ref{eqVR})
are then used to provide a graphical form for the potentials $V_{L}$ and $V_{R}$. We
applied this procedure for the $a=1/3$ case, whose results are depicted in
Figs. \ref{figV_explic}a-b. Such graphics reveal that both procedures
described here lead to almost the same results.

\begin{figure}[tbp]
{\includegraphics[{angle=0,width=7cm,height=6cm}]{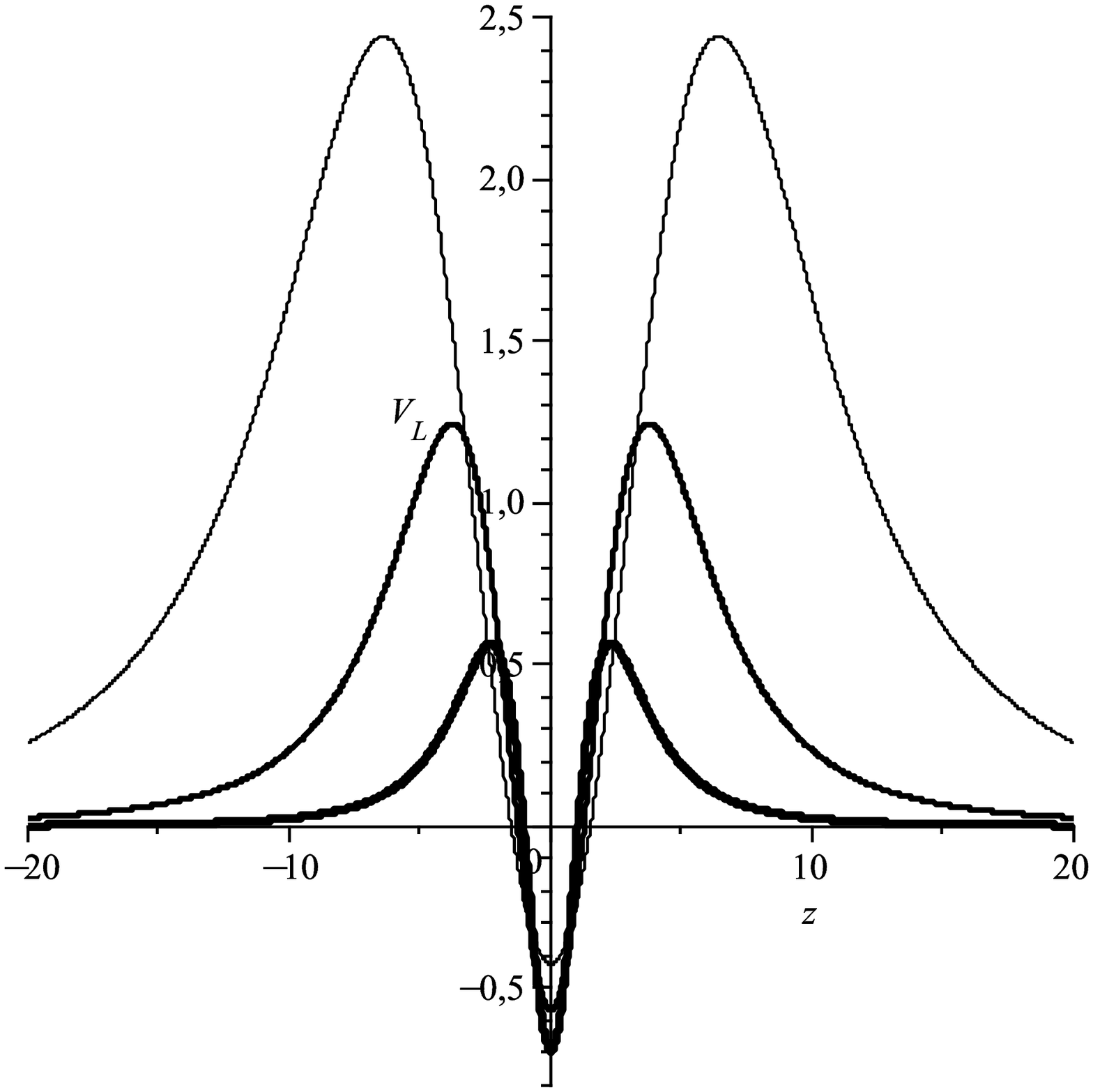}} {%
\includegraphics[{angle=0,width=7cm,height=6cm}]{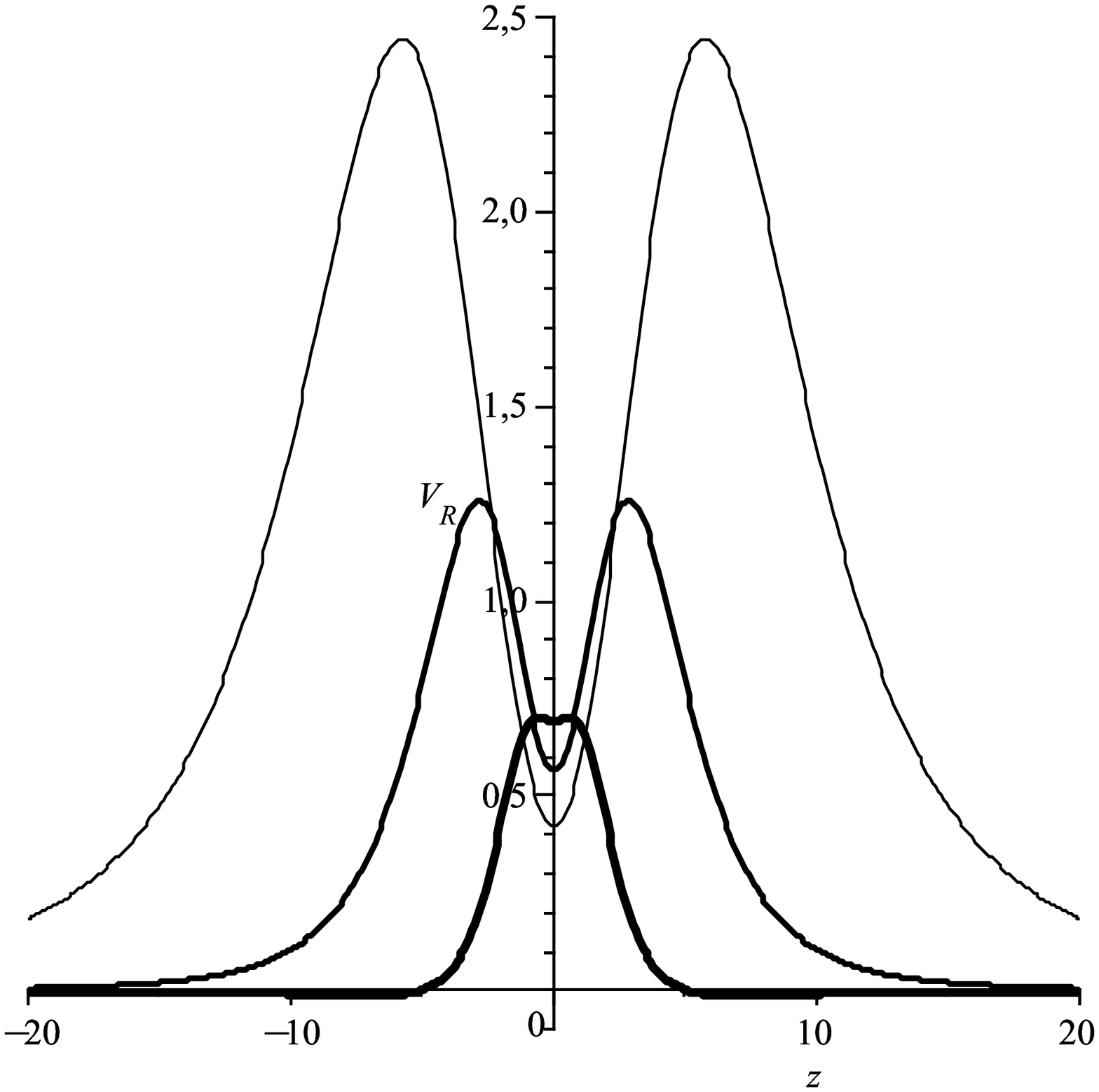}}
\caption{Schr\"odinger-like potentials (a) $V_L(z)$ (left) and (b) $V_R(z)$
(right) for $a=0.05$ (thinner trace), $a=0.10$ and $a=0.20$ (thicker trace).}
\label{figVL}
\end{figure}

Now, we can use the general numerical procedure to investigate the potential
$V_{L}$ for other values of $a$. Some results are exhibited  in Fig. \ref%
{figVL}a. We note from this figure that $V_{L}$ has a volcano-like shape,
that goes asymptotically to zero far from the brane. This form of potential
is usually found in problems involving gravity localization. Here, we can
search for the existence of a zero mode followed by a continuum of massive
modes by means of an analogy. Indeed, we note that the potential is negative
at the brane location; this guarantees the existence of a normalized zero
mode after integrating Eq. (\ref{HL}) for $m=0$:
\begin{equation}
L_{0}\propto \exp \bigl(-\eta \int_{0}^{z}dz^{\prime }e^{A(z^{\prime })}\phi
(z^{\prime })\chi (z^{\prime })\bigr).
\end{equation}%
Fig. \ref{figVL}b shows that the potential $V_{R}$ is always positive at the
brane position, which is not compatible with the existence
of bound fermions with right chirality. However, we note the appearance of a
hole in the potential that grows for lower values of $a$. This fact is new
and our hypothesis is that it can be related, in some way, to the internal
structure \cite{bg} of the brane constructed with two fields. It is
remarkable that such behavior of the $V_{R}$ potential was not observed in a
previous treatment for one-field models. The appearance of this hole in the
potential can be responsible for resonances or at least for a light increase
in the decay rate of massive fermions on the brane. Note also that the hole
around $z=0$ for $V_{R}$ is absent for larger values of $a $, in principle
prohibiting the existence of resonances for this range of parameter. In order to look for
resonant effects, we must consider now the massive modes, solving
numerically the Schr\"{o}dinger-like equation with purely numeric potentials.

\subsection{Right-handed fermions}

Firstly we will consider the case of right-handed fermions with the small
parameter $a=0.05$. We can vary $m^2$ to get the wavefunctions $R_m(z)$. Eq.
(\ref{VschR}) shows that we need two initial conditions. As we are dealing
with thick brane with no tension at $z=0$, the Israel junction conditions
applied to the brane give $R(z)$ and $R^{\prime}(z)$ continuous at $z=0$.
Note that the absence of negative values for the potential $V_R$ prohibits
the existence of bound states. However, the tunneling process across the
barrier (localized at the region between the maximum of the potential and the origin) can result
in different rates for the leaking process depending on the mass of the
modes. Here the region of interest corresponds to masses satisfying $%
0.42<m^2<2.44$, where the tunneling process is effective (see Fig. \ref%
{figVL}b - thinner trace).

\begin{figure}[tbp]
\includegraphics[{angle=0,width=7cm,height=6cm}]{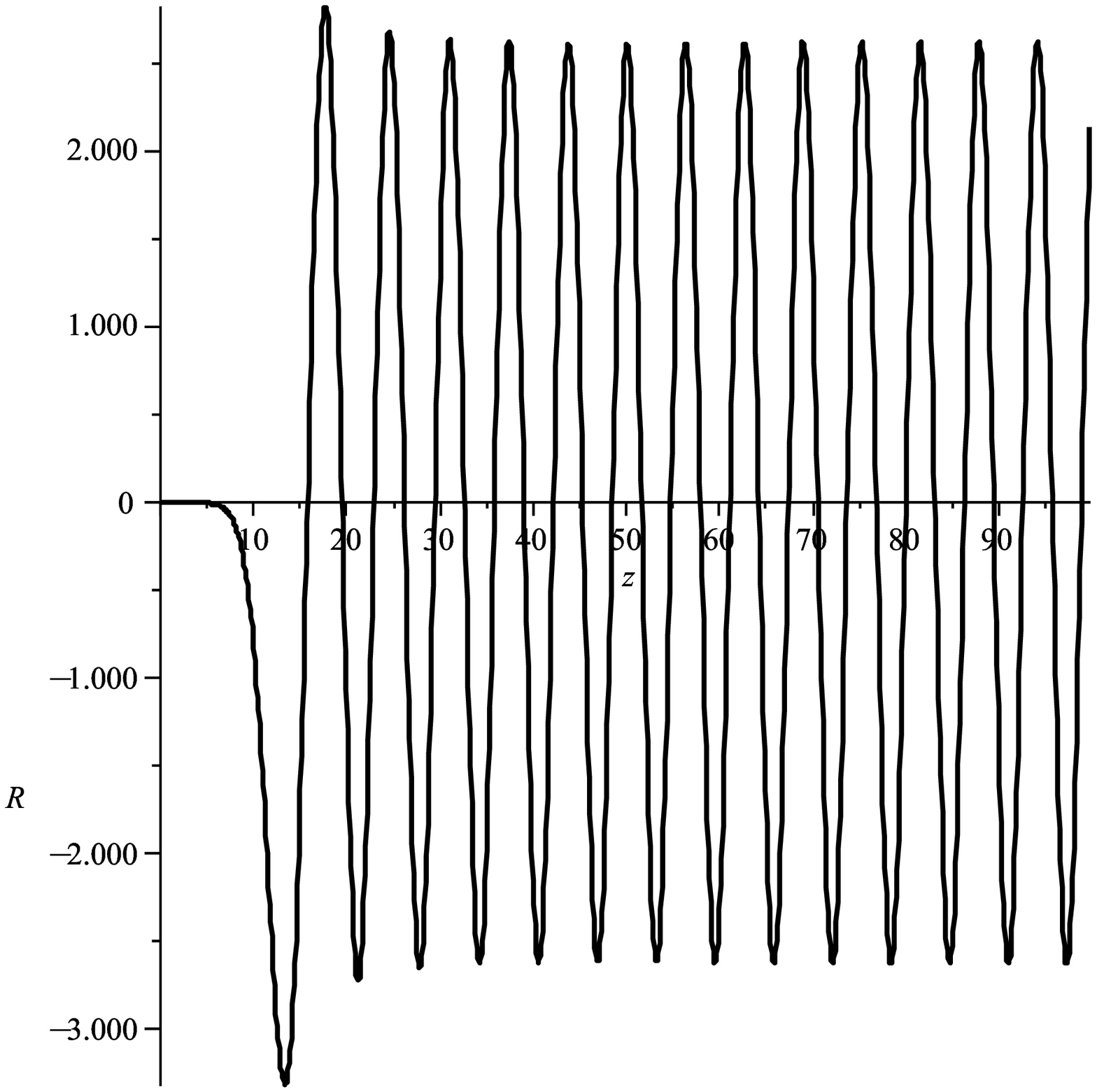} %
\includegraphics[{angle=0,width=7cm,height=6cm}]{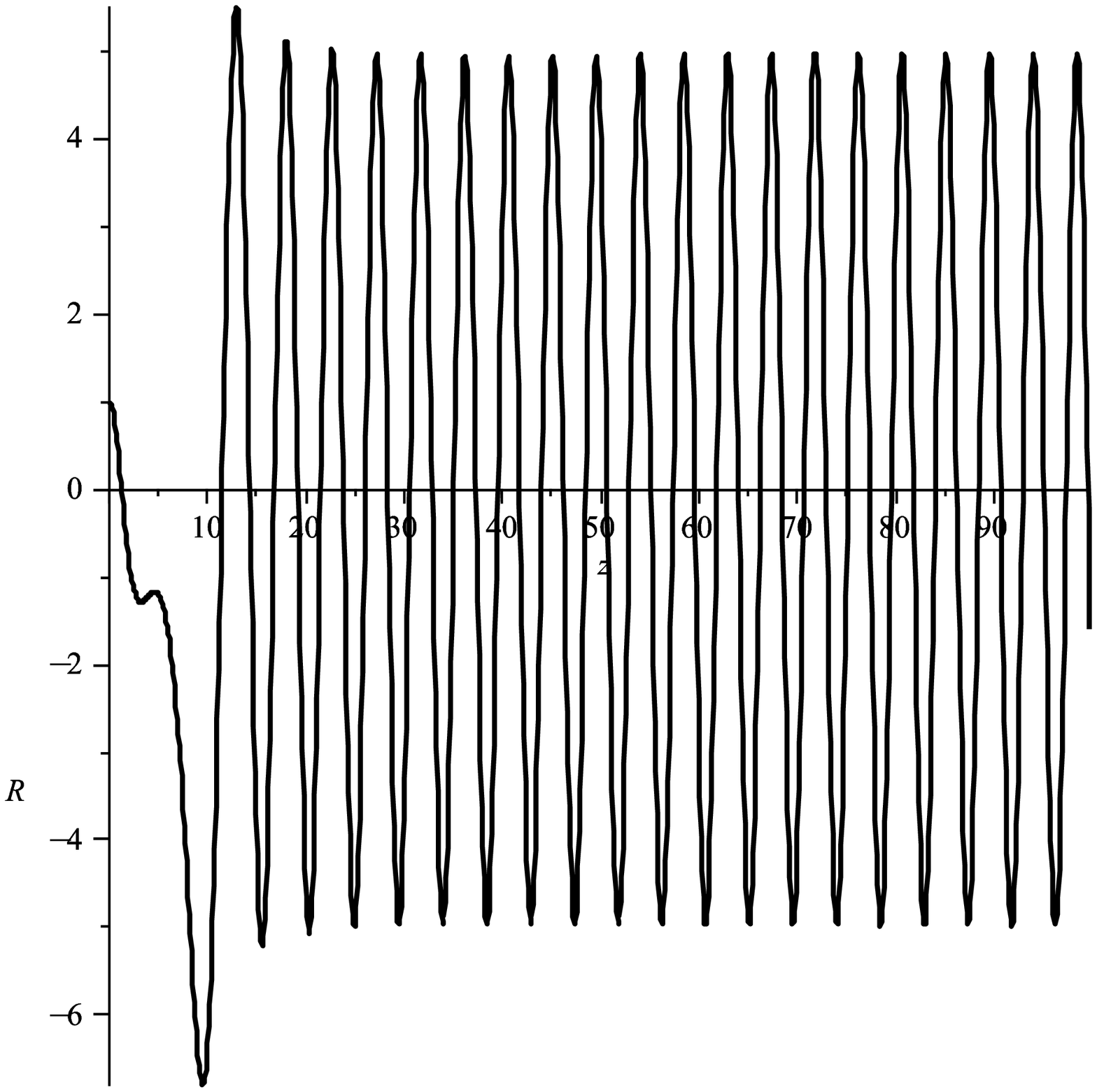}\newline
\includegraphics[{angle=0,width=7cm,height=6cm}]{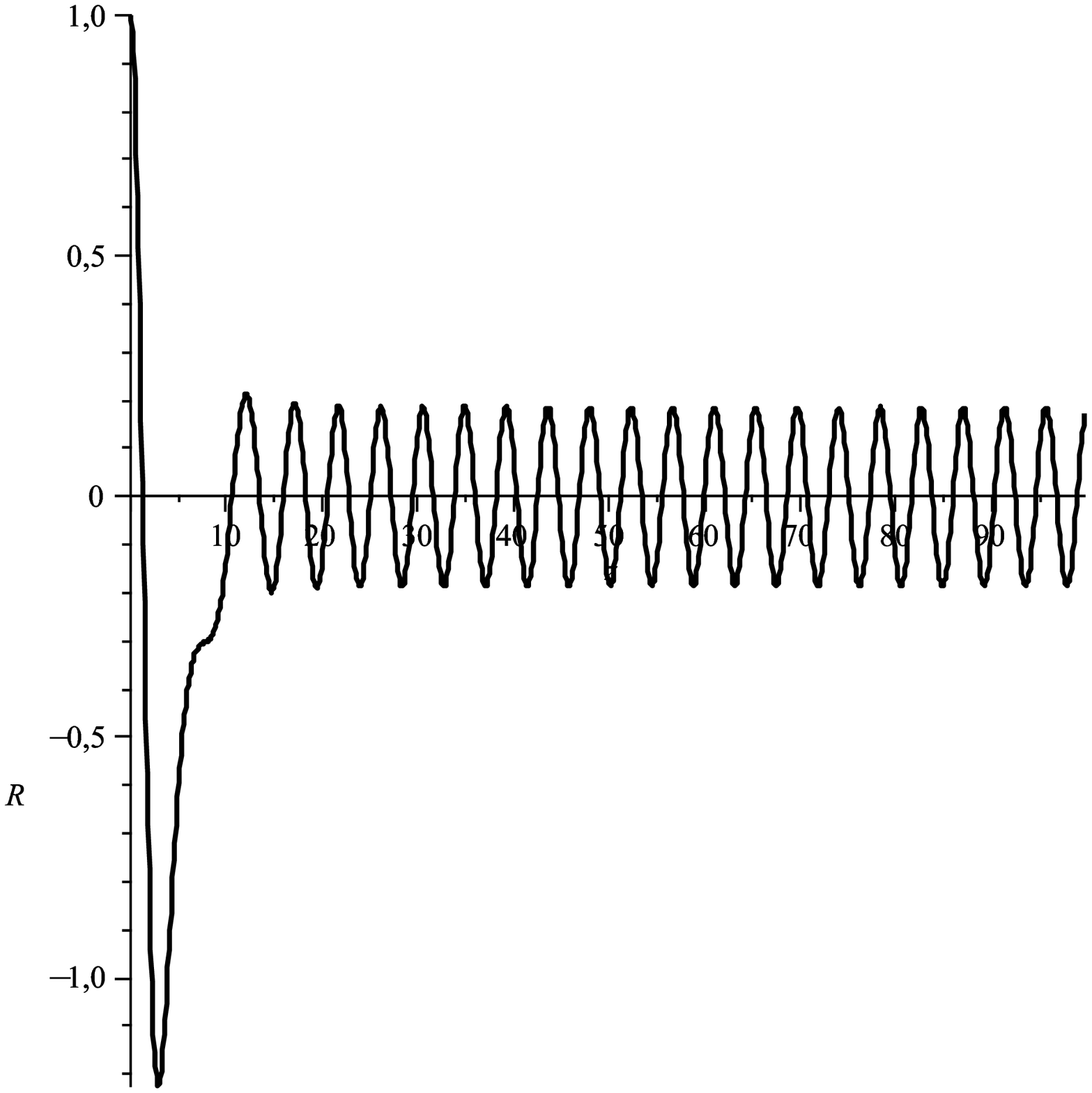} %
\includegraphics[{angle=0,width=7cm,height=6cm}]{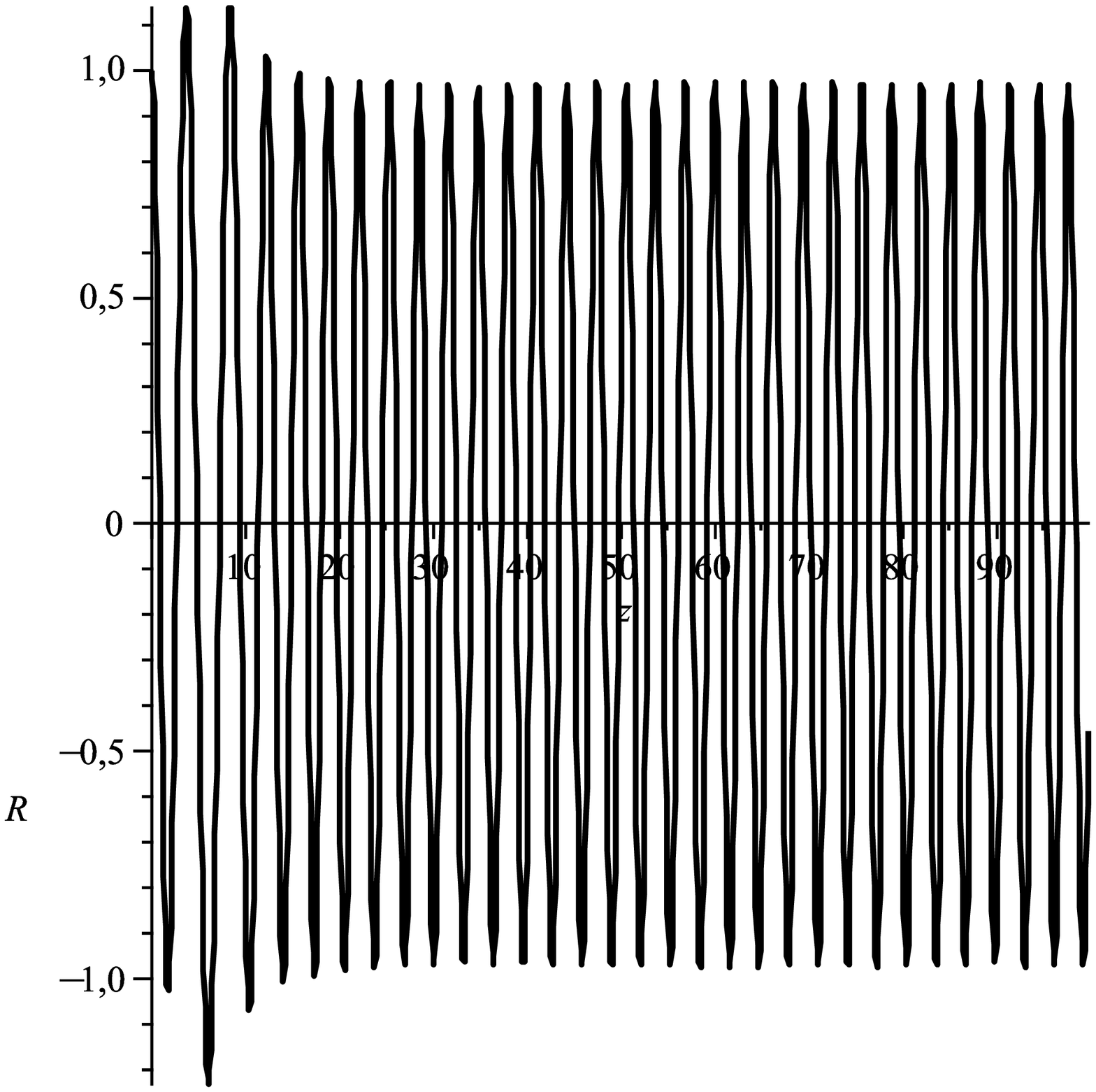}
\caption{Non-normalized even massive right-handed fermionic modes for $a=0.05
$ with (a) $m^2=1$ (upper left), (b) $m^2=2$ (upper right), (c) $m^2=2.1$
(lower left) and (d) $m^2=4$ (lower right).}
\label{figVRa005}
\end{figure}

Figs. \ref{figVRa005}a-d shows typical even massive fermionic modes before
normalization for the region $z>0$. Figs. \ref{figVRa005}a-c belongs to the
more interesting region $0.42<m^2<2.44$ whereas Fig. \ref{figVRa005}d is for
a higher massive mode. The lower massive modes show that there is a
transient behavior followed by typical plane wave oscillations,
characteristic of a free mode. This shows that these modes represent massive
fermions that certainly will be leaked from the brane. Comparing Fig. \ref%
{figVRa005}a with Figs. \ref{figVRa005}b and \ref{figVRa005}c, we note that
the transition region roughly corresponds to $-20<z<20$. Note also that for
lower values of $m^2$ the values of the non-normalized wavefunctions $R(z)$
depart largely from the initial value 1. More important is to estimate how
larger is the solution $R(z)$ in the transition region $-20<z<20$ in
comparison to the amplitude of the plane wave oscillations. Indeed, one
would expect that broader differences would correspond to greater lifetimes
for the massive fermionic states near to the brane. This difference assumes
a huge amount for $m^2=2.1$, as we can see from Fig. \ref{figVRa005}c.

This lead us to the point of normalization of the massive modes and to the
better estimate of $R_m(0)$. Their importance is that we can
rewrite Eq. (\ref{VschR}) as $\mathcal{O}_R^{\dagger}\mathcal{O}%
_RR(z)=m^2R(z)$, which ensures that we can interpret $|R_m(0)|^2$ as the
probability for finding the massive modes on the brane. In particular, large
peaks in the distribution of $R(0)$ as a function of $m$ would reveal the
existence of resonant states closely related to the existence of a brane
with internal structure.

\begin{figure}[tbp]
\includegraphics[{angle=0,width=7cm,height=6cm}]{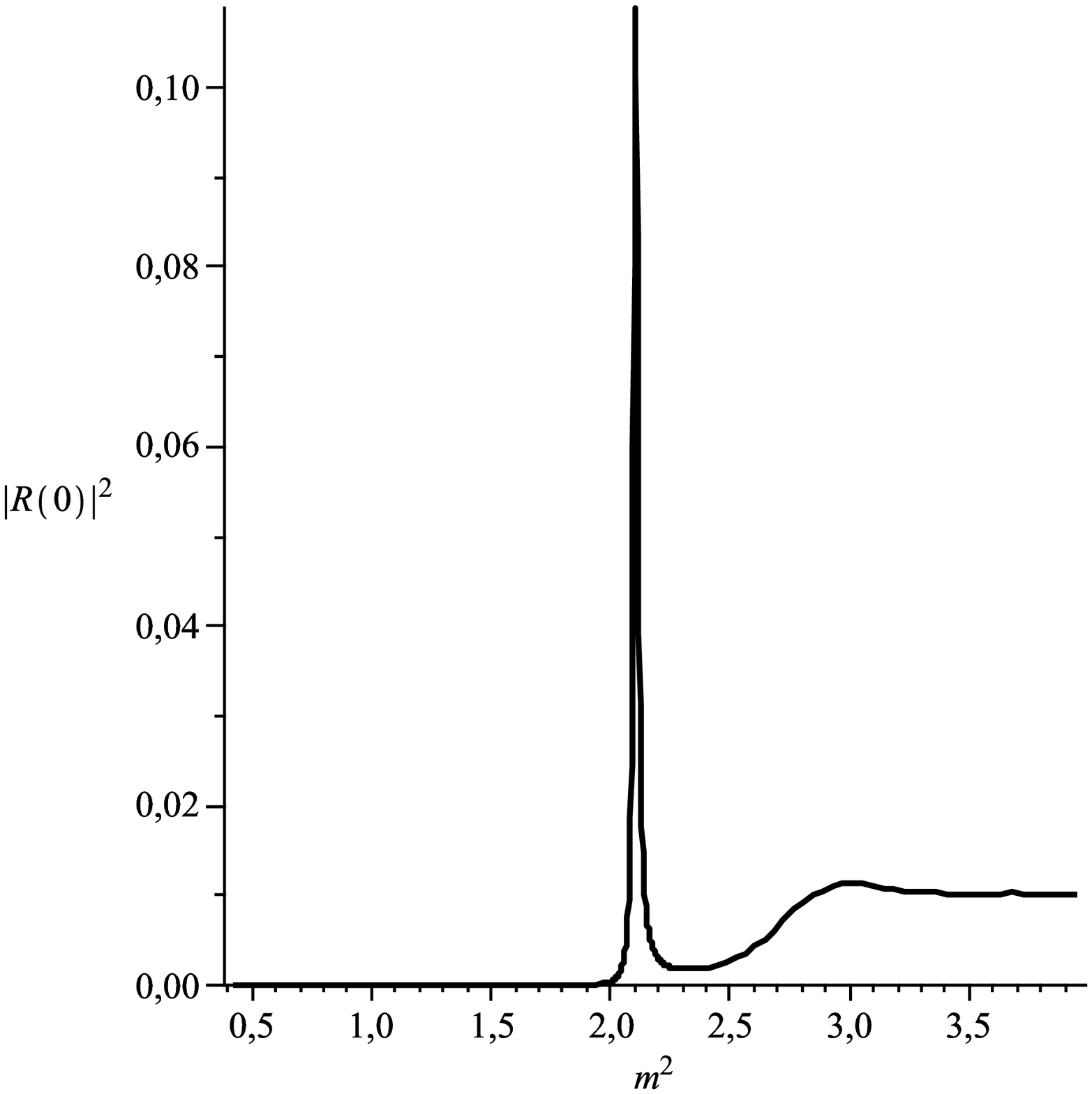} %
\includegraphics[{angle=0,width=7cm,height=6cm}]{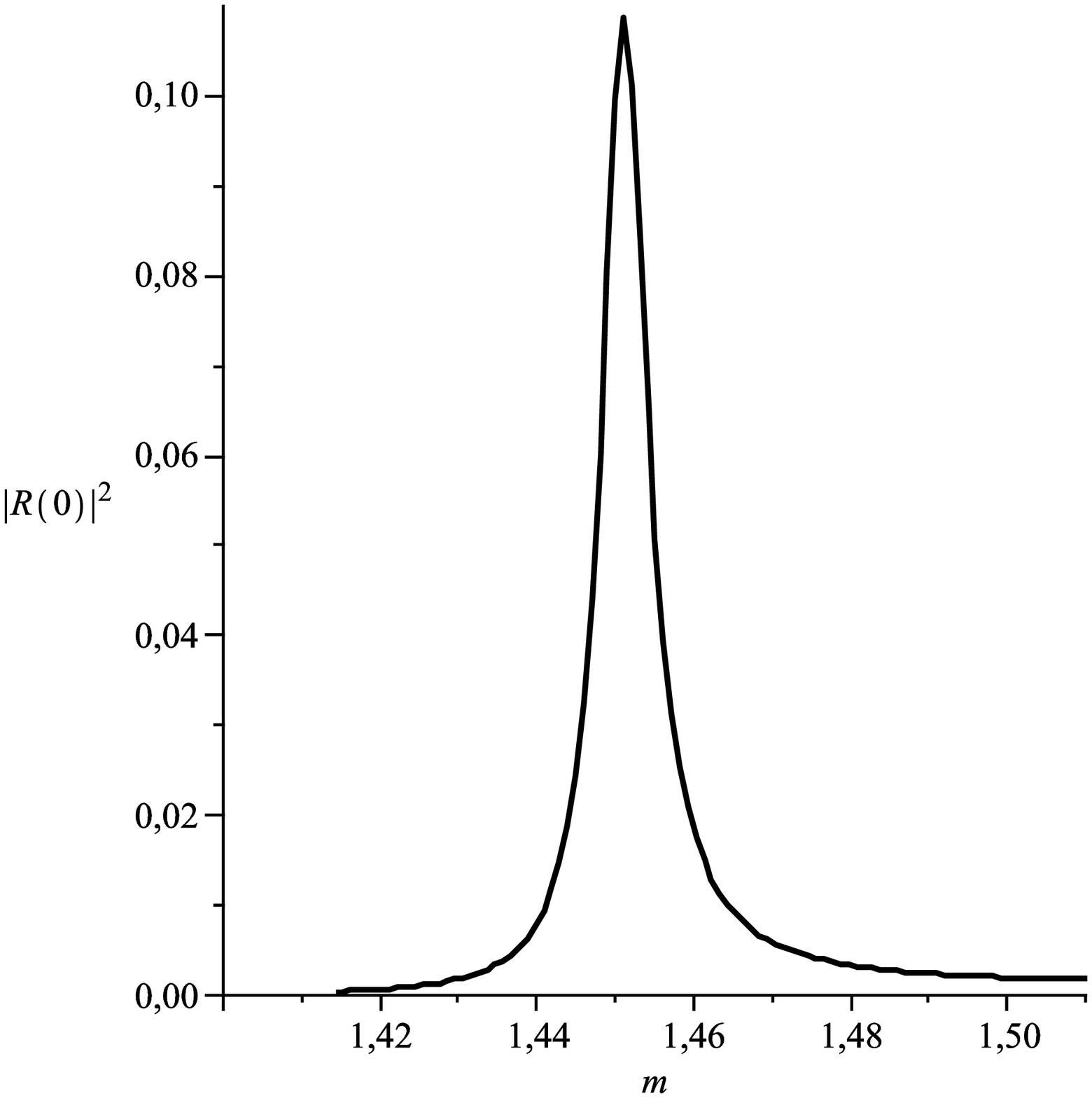}
\caption{(a) Normalized squared wavefunction for right-handed fermion on the
brane, $|R(0)|^2 $, as a function of $m^2$. Note the high peak
characteristic of resonance. For the present case we have $a=0.05$. (b)
Normalized squared wavefunction for right-handed fermion on the brane, $%
|R(0)|^2 $, as a function of $m$, identifying better the region defining the
peak of resonance.}
\label{figRsq}
\end{figure}

We can consider a normalization procedure for the wavefunctions $R_m(z)$ in
a box with borders located far from the turning point, where the solutions
have characteristics of plane waves toward the brane. In this way, we
considered a box with $-100<z<100$. An important information can be obtained
from the normalized $|R_m(0)|^2$, since from it we can compare the relative
probability for finding a massive mode on the brane. The result is depicted
in Fig. \ref{figRsq}a. We note from the figure a huge peak around $m^2=2.1$,
characterizing the occurrence of a resonance and a long-lived massive
fermionic mode on the brane. We note from the figure that low massive modes
have negligible associated probabilities, whereas high massive modes are
characterized by a plateau in the probability distribution. This is related
to the finite size of the box used for normalization. Indeed, for $m^2>>V_R$
we can approximate the Schr\"odinger equation as $-R^{\prime\prime}(z)\sim
m^2R(z)$, with normalized solution $R(z)\sim1/\sqrt{z_{max}}\cos(mz)$, where
$2z_{max}$ is the size of the box used for normalization. For our choice $%
z_{max}=100$, we have $|R(0)|^2=1/(z_{max})=0.01$, corresponding to the
plateau observed in Fig. \ref{figRsq}a.

Since in our model the extra dimension is infinite, one may wonder
about the influence of $z_{max}$ on the normalization of the spectral
density. Indeed, another choice of $z_{max}$ would lead to another
plateau in the probability distribution. The physical
information, however, is contained in the value of the resonance peak, which
does not depends on $z_{max}$ since it is chosen sufficiently large.

Fig. \ref{figRsq} shows the peak of resonance with the horizontal scale now
taken as $m$. We can estimate the life-time $\tau$ for the observed
resonance from the width at half maximum of the peak appearing in the
figure. Indeed, the width in mass $\Gamma=\Delta m$ of the peak is related
to the life-time. In this way, in its own reference frame, the fermion
disappears toward the extra dimension with time scale $\tau\sim\Gamma^{-1}$
\cite{grs}. The peak was located at $m=1.451(1)$ with the maximum
probability around $0.109$. The width at half maximum is $\Delta m=0.0083$,
resulting in a lifetime $\tau_a=\tau_{0.05}=120$. These results are in
arbitrary units, and their significance will be better understood after
comparison with the results for other values of the parameter $a$.

\begin{figure}[tbp]
{\includegraphics[{angle=0,width=7cm,height=6cm}]{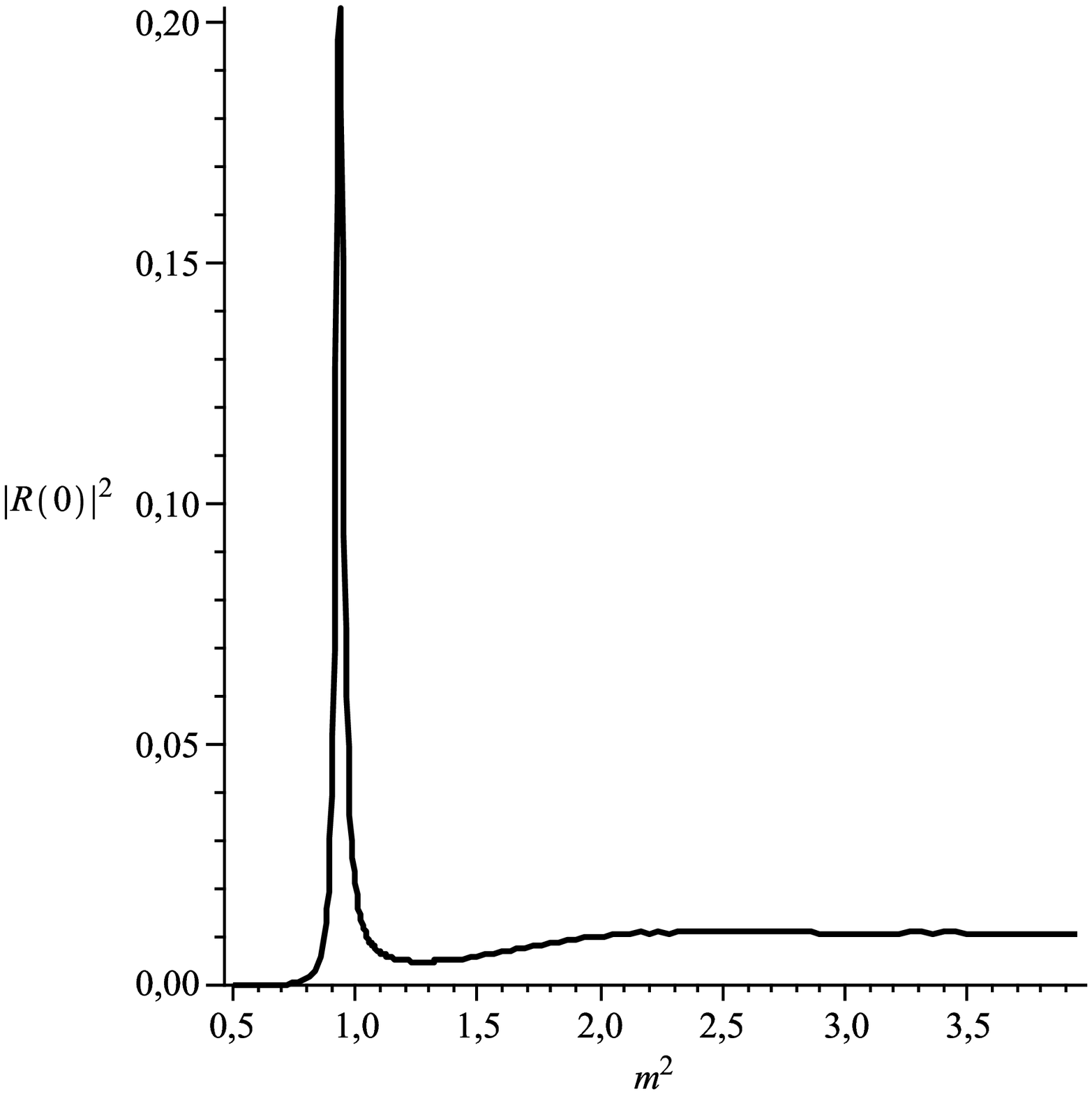}} {%
\includegraphics[{angle=0,width=7cm,height=6cm}]{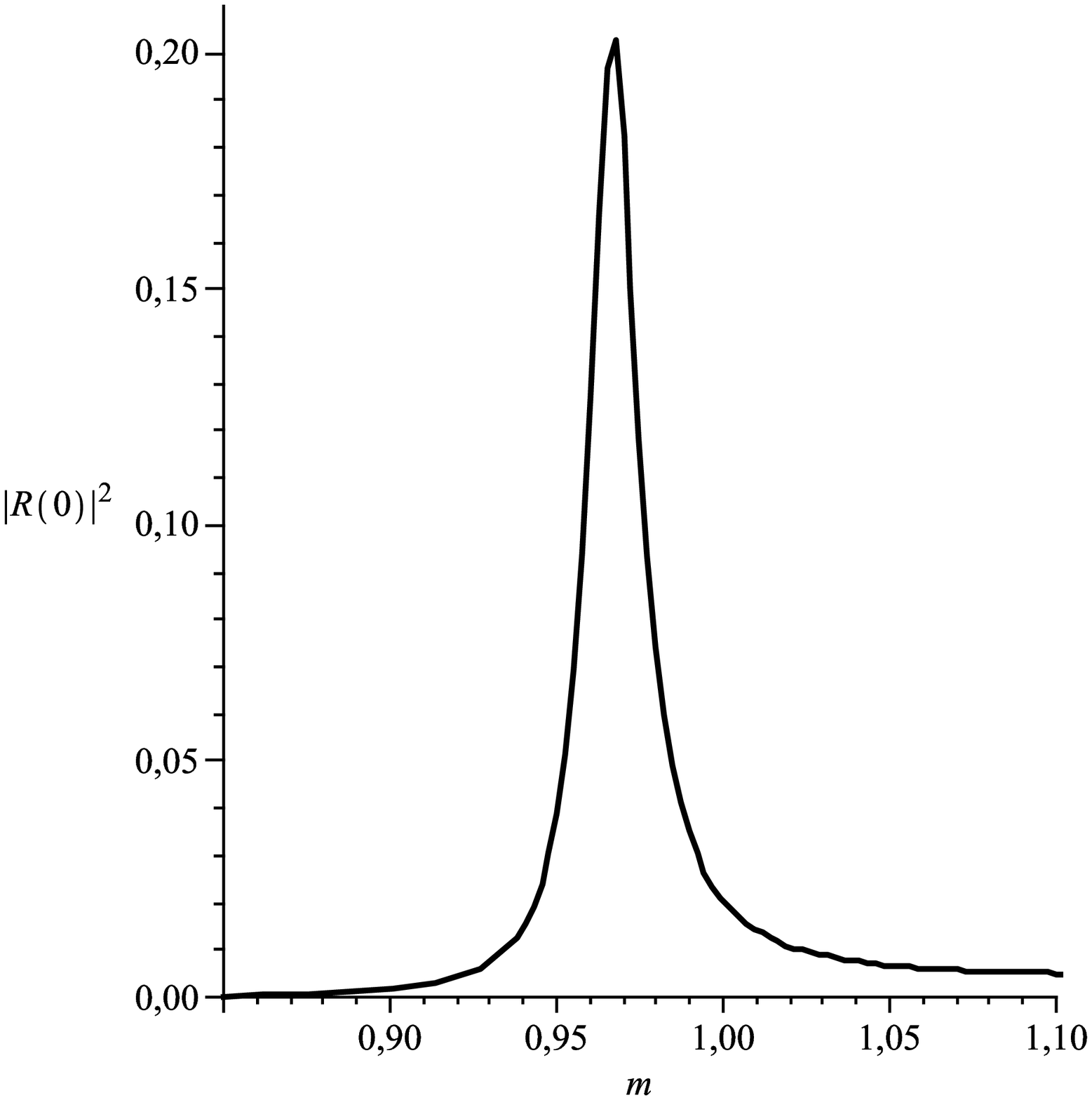}}
\caption{(a) Normalized squared wavefunction for right-handed fermion on the
brane, $|R(0)|^2 $, as a function of $m^2$, for $a=0.10$ and $-100<z<100$.
(b) Normalized squared wavefunction for right-handed fermion on the brane, $%
|R(0)|^2 $, as a function of $m$, identifying better the region defining the
peak of resonance.}
\label{figVRa010}
\end{figure}

Now, we repeat our analysis for other values of $a$ to investigate the
effects of the coupling parameter $a$ between the fields $\phi$ and $\chi$
for the existence of resonances and their lifetimes.

For $a=0.10$ the results for $|R(0)|^2 $, as a function of $m^2$,
are depicted in Fig. \ref{figVRa010}a. Note from the figure that we also
have one resonant peak as before. Now, however, the larger value of $a$
results in a smaller value of the mass for the resonant state. Also, the
resonant peak is more pronounced. Fig. \ref{figVRa010}b shows that the peak
was located at $m=0.968$ with the maximum probability around $0.202$. The
width at half maximum is $\Delta m=0.02$, resulting in a lifetime $%
\tau_a=\tau_{0.1}=50$. Comparing this result with $\tau_a=\tau_{0.05}=120$,
we conclude that smaller values of $a$ result in resonances with larger
lifetimes.

\begin{figure}[tbp]
{\includegraphics[{angle=0,width=9cm}]{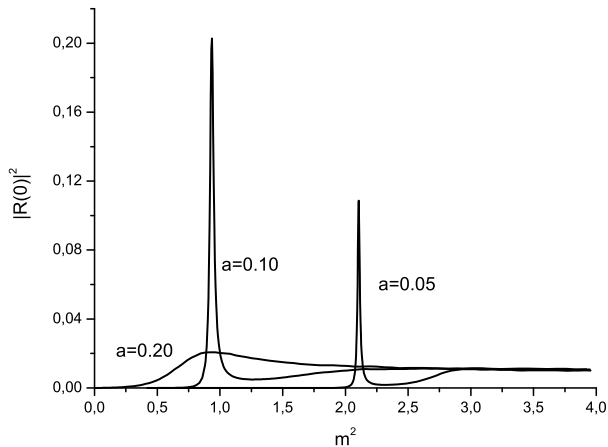}}
\caption{ Normalized squared wavefunction for right-handed fermion on the
brane, $|R(0)|^2 $, as a function of $m^2$. We used $-100<z<100$, and the
resonant peaks correspond to $a=0.05$ (third narrow and smaller peak), $%
a=0.10$ (higher peak). Also it is showed the broader peak for $a=0.20$.}
\label{figresonances}
\end{figure}

We compare both cases in Fig. \ref{figresonances}, where we display the
resonances for $a=0.05$ (second smaller peak) and $a=0.10$ (first larger
peak). We know that smaller values of $a$ correspond to branes with richer
internal structure \cite{bg2}. Then the observed thinner peaks for smaller
parameters $a$ means resonances more pronounced and a trapping mechanism
more effective. This is connected with the larger amount of matter now
forming the brane due to the coupling of the scalar field $\phi$ to the
other field $\chi$. 

\begin{figure}[tbp]
{\includegraphics[{angle=0,width=7cm,height=6cm}]{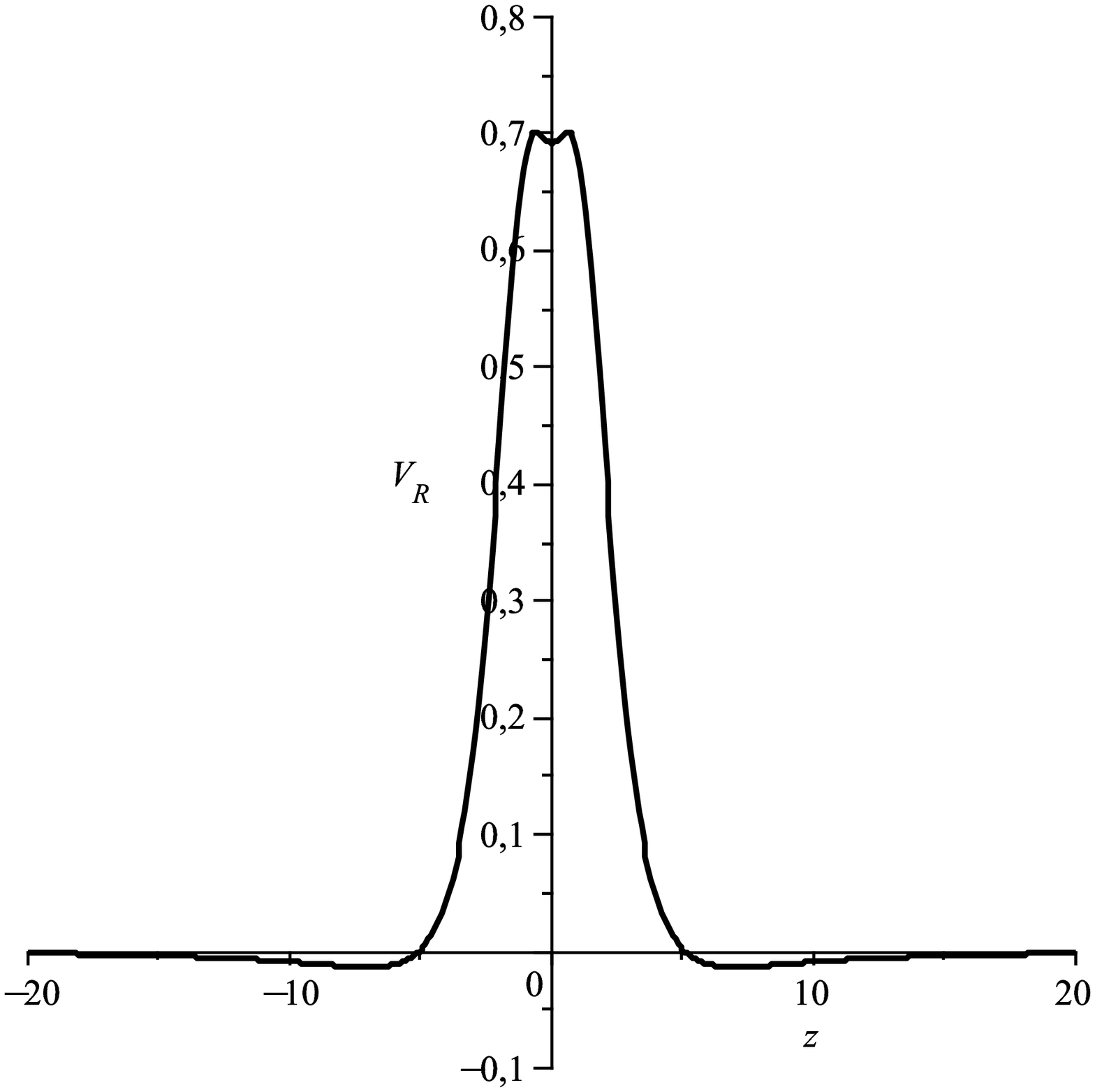}} {%
\includegraphics[{angle=0,width=7cm,height=6cm}]{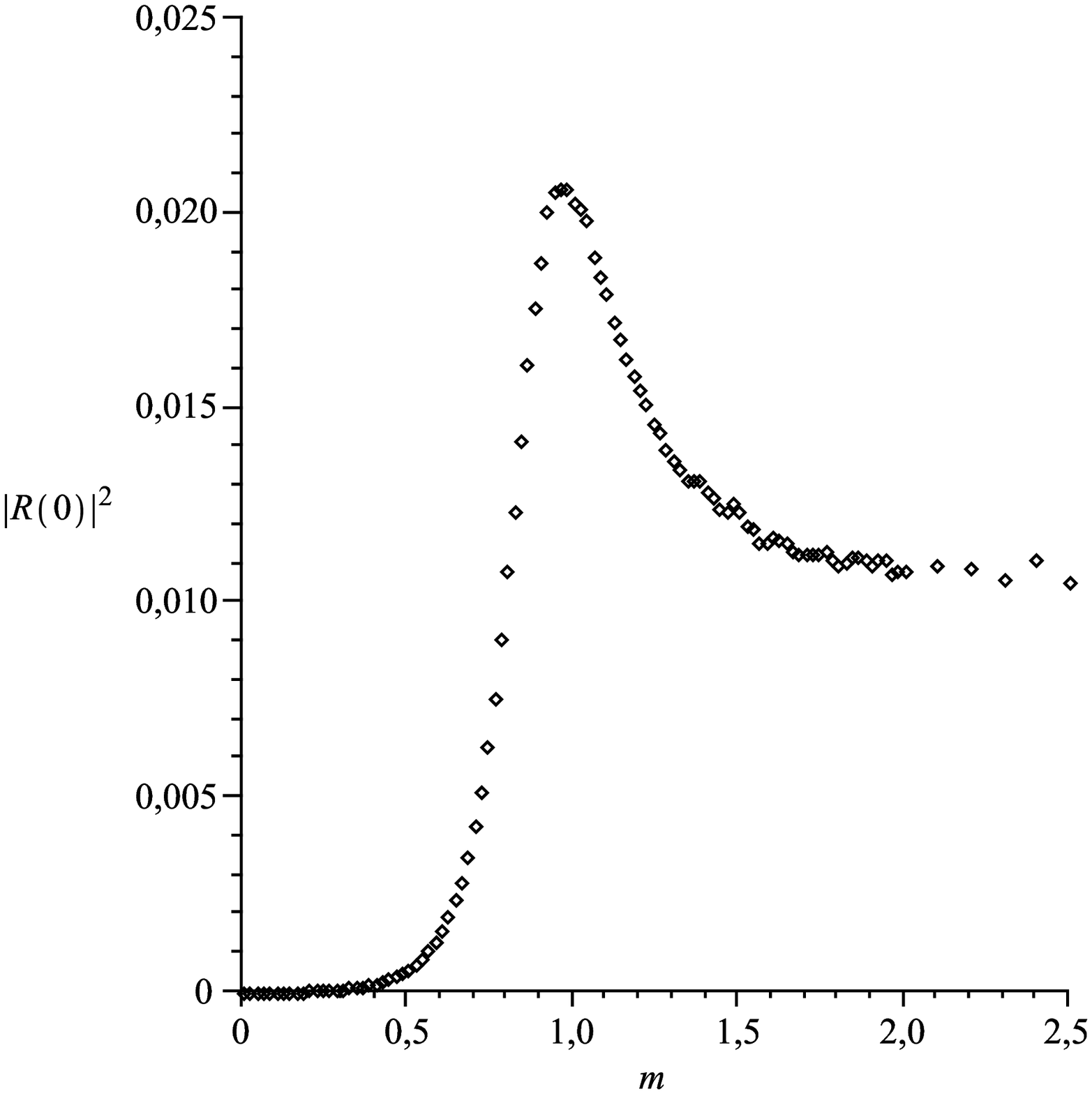}}
\caption{(a) Schr\"odinger-like potential $V_R(z)$ for $a=0.20$. Note the
appearance of a region with negative potential, absent for the lower values
of $a$ considered previously. (b) Normalized squared wavefunction for
right-handed fermion on the brane, $|R(0)|^2 $, as a function of $m$. Note
now the presence of a broader peak with $\Delta m>m$.}
\label{VRa020}
\end{figure}
The reduction of the lifetime for resonances with larger values of $a$ poses
the question if there are no resonances at all on the brane for values of $a$
above a certain threshold $a^*$. We noted the decreasing of the associated
mass for the resonances with larger values of $a$. Also, it was noted a
qualitative change on the potential $V_R$ above a certain value $a^*$. For
example, for $a=0.20$, Fig. \ref{VRa020}a shows that now the potential
assumes the form of a volcano. One
characteristic of this type of potential is the occurrence of a zero-mode
when the central part, around the brane, assumes negative values. This is
the potential observed in Fig. \ref{figVL}a for left-fermions. Here,
however, the central part of the potential is positive and the zero-mode is
absent. For $a=0.20$ there were observed no resonances on the brane.
In fact, Fig. \ref{VRa020}b depicts the probability $|R(0)|^2 $ as a
function of the mass of excitations. Here we can see that the width at half
maximum $\Delta m$ is larger that the position of the peak $m=0.95(1)$. This
characterizes a state with a lifetime $\tau=1/{\Delta m}$ too small to
result in any physical effect, characterizing the absence of a resonant
state. 

\begin{figure}[tbp]
{\includegraphics[{angle=0,width=7cm}]{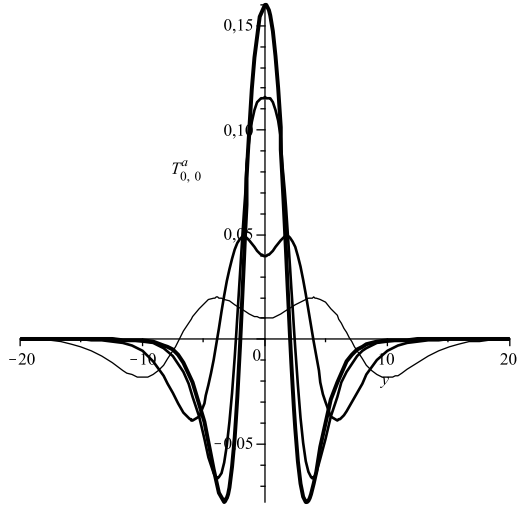}} 
\caption{(a) Plots of the matter energy density $T^{\,0.05}_{00}(y)$
(thinner trace), $T^{\,0.10}_{00}(y)$,$T^{\,0.17}_{00}(y)$ and $%
T^{\,0.20}_{00}(y)$ (thicker trace).}
\label{figT00}
\end{figure}

We can relate these results with the behavior of the energy density of the
brane without fermions as studied in Ref. \cite{bg2}. The energy density $%
T_{00}^{\,a}(y)$ as a function of the coordinate $y$ and for a fixed
parameter $a$ is given by
\begin{eqnarray}  \label{T00}
T_{00}^{\,a}(y)\!\!=\!\!e^{2A(r)}\biggl[\frac12\biggl[ \bigg(\frac{d\phi(y)}{%
dy}\biggr)^2+ \bigg(\frac{d\chi(y)}{dy}\biggr)^2\biggr] + V(\phi(y),\chi(y)) %
\biggr].
\end{eqnarray}
Fig. \ref{figT00} shows that the structure of the brane changes when the
parameter is $a= 0.17$. Indeed, we note that for $0.17<a<0.5$ the energy
density characterizes a defect with the energy density centered around a
central peak, whereas for $a<0.17$ the appearance of two symmetric peaks
signals the occurrence of a brane with internal structure (more details in
Ref. \cite{bg2}). Now, we connect this known result with our findings of
resonance peaks for small values of $a$ and with the absence of such
resonances for larger values of $a$. There appears to be a connection
between the occurrence of metastable states on the brane and the occurrence
of two peaks in the energy density characterizing these branes. From this we
conclude that branes with internal structure favor the appearance of
resonance states for right-handed fermions. Also, for branes without
internal structure, we were not able to find resonance states.

A further point to remark is that we have considered analytical fittings for
the Schr\"odin\-ger potentials $V_R(z)$ in terms of finite series with
oscillating terms. For fittings that agreed visually with our numerical
points, the Runge-Kutta method was applied and the normalized massive modes
were constructed. The results for the resonances were roughly the same,
corroborating our conclusions.

\subsection{Left-handed fermions}

For left-handed fermions, as we already noted, there is a normalized zero
mode. We repeated the analysis from the previous section and also found
resonances. However, for left-handed fermions there are differences with
respect to the position of the resonance peak, in comparison to which was
obtained for right-handed ones. For instance, for $a=0.05$, we found a
resonance peak around $m=1.59$, with $\Delta m=0.07$, corresponding to a
life-time $\tau=14$. When comparing with corresponding results for
right-handed fermions ($m=1.45$ and $\tau=120$), we see that the left-handed
resonance is the more massive and the less stable. Analysis for the
resonance peak for $a=0.10$ showed that, similarly to the observed for
right-handed fermions, an increasing $a$ also thickens the resonance
peak for left-handed fermions.

\subsection{Correspondence between the spectra and realization of Dirac
fermions}

The lack of correspondence between the spectra of left- and
right-handed fermions appears to be in contradiction to the fact that $V_L$ and $V_R$ are superpartner potentials. The reason for this is that the parity-odd modes
do not couple to the four dimensional braneworld at $z=0$, because for such
modes we have $R(0)=0$ and $L(0)=0$. This was also noted in the case of
gravity localization in another model \cite{llatas}. After we submitted this
work, we knowed about Ref. \cite{liu}, where a study of odd parity modes is
done. According to ref. \cite{liu}, we set $L(0)=0$ and $L^{\prime}(0)=5$
for parity odd wavefunctions as our starting point for the application of
the numerical Numerov method. We use the scheme with step $h$ in $z$
corresponding to the points $z_i$ described by
\begin{eqnarray}
L(0)&=&0\,, \\
L(h)&=&hL^{\prime }(0)\,, \\
L(z_i)&=&\frac{2[1+5F(z_{i-1}]L(z_{i-1})-[1-F(z_{i-2})]L(z_{i-2})}{1-F(z_i)}\,,
\end{eqnarray}
where
\begin{equation}
F(z)=-\frac 1 {12}h^2(V_L(z)-m^2)\,.
\end{equation}
The condition $L^{\prime}(0)=5$ is arbitrary and will be fixed after the
normalization process.

Thick branes are extended objects along the extra dimension that
allow us to interpret, after normalizing $L(z)$, the probability for finding
the massive modes on the brane (not necessarily at $z=0$) as $%
\int_{-z_b}^{z_b}dz|L_m(z)|^2$. The parameter $z_b$ is chosen in order to
allow the influence of the odd modes in small regions around $z=0$. We can
consider a normalization procedure for the wavefunctions $L_m(z)$ in a box
with borders located at $z=\pm z_{max}$, far from the turning point, where
the solutions have characteristics of plane waves far from the brane. After
normalization of $L(z)$, the probability for finding the massive modes on
the brane is given by \cite{liu}
\begin{equation}
P_{brane}=\frac{\int_{-z_b}^{z_b}dz|L_m(z)|^2}{%
\int_{-z_{max}}^{z_{max}}dz|L_m(z)|^2}.
\end{equation}
\begin{figure}[tbp]
{\includegraphics[{angle=0,width=7cm}]{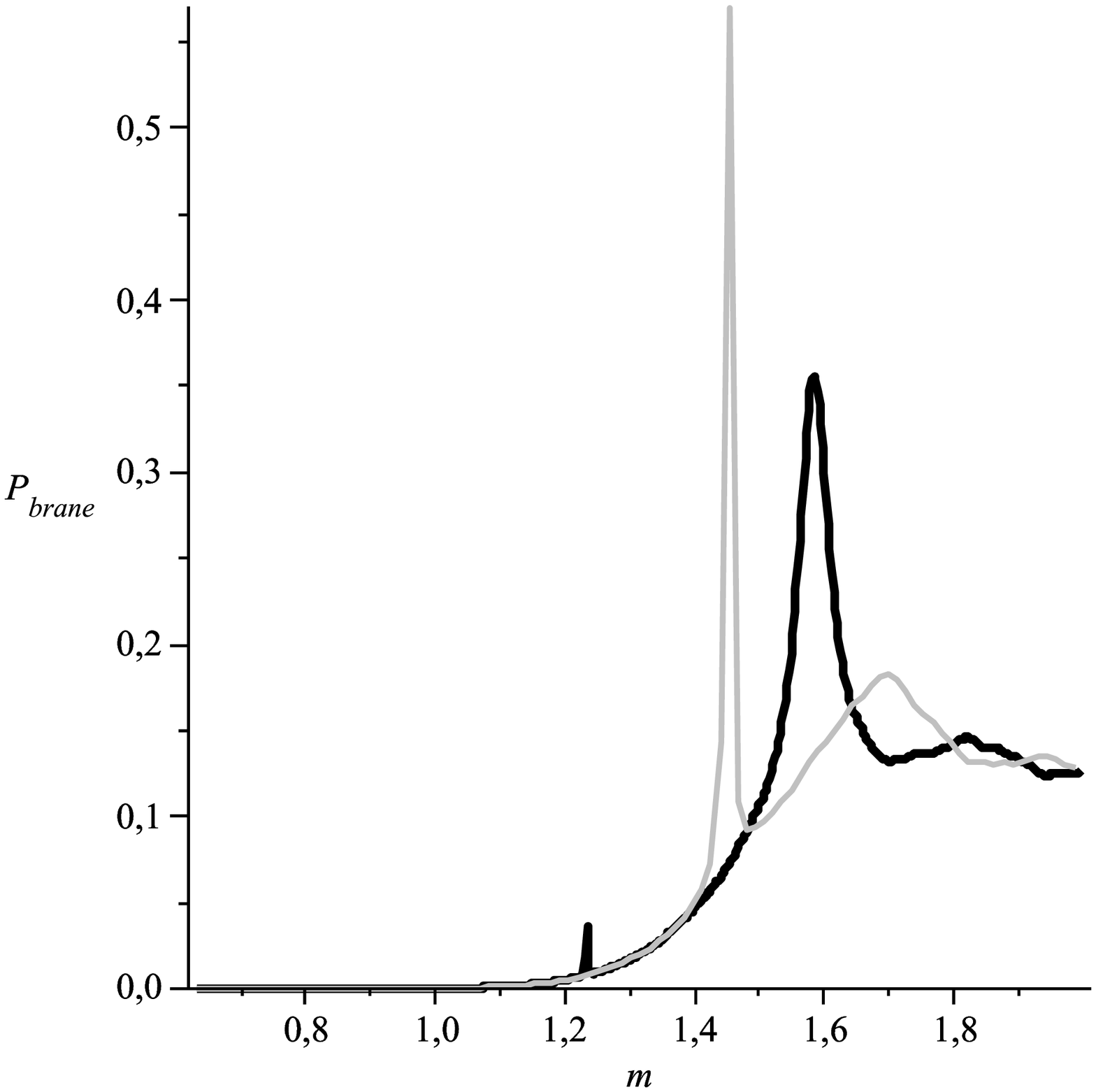}} 
{\includegraphics[{angle=0,width=7cm}]{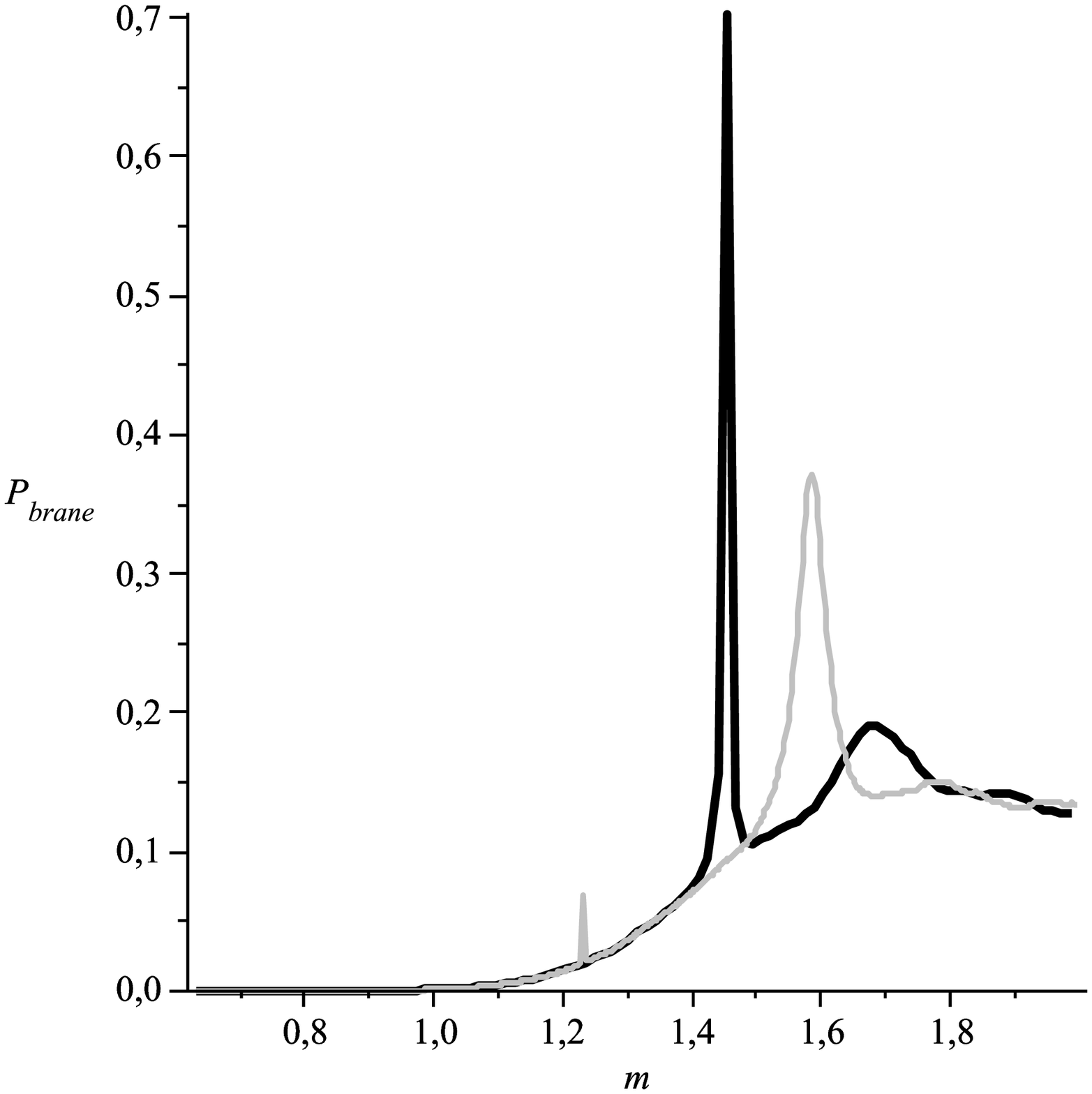}} 
\caption{Plots of $P_{brane}$ for $z_{max}=100$, $z_b=10$, $a=0.05$. Curves
for a) $V_R$ (left) and b) $V_L$ (right) constructed with even (thicker trace) and odd
(thinner gray trace) wavefunctions. }
\label{figLR_odd}
\end{figure}
We followed the same procedure for right-handed fermions and depicted in
Fig. \ref{figLR_odd} the results for $P_{brane}$ in the particular case of $%
a=0.05$. The values of $m$ for the resonance peaks in the figure
corresponding to the parity even wavefuntions agree with the obtained
previously in the plots of the normalized $|R(0)|^2$ (Fig. \ref{figRsq}b)
and $|L(0)|^2$. The figure also shows that with this procedure the resonance
peaks for left- and right-quirality agree with good precision and the
formation of Dirac fermions is realized. Larger values of the parameter $z_b$
do not change the results for the position of the resonance peaks, but small
values of $z_b$ are more efficient to identify the peaks. One example of
such procedure in a model with three thin branes was done in Ref. \cite{m},
were one of the wavefunctions (right or left-handed) is symmetric and the other
corresponding to the same mass is antisymmetric with respect to the center of the orbifold. This is a
consequence of demanding the action to be even under the $Z_2$ orbifold
symmetry \cite{gross}. In this way it is shown the existence of metastable
massive Dirac modes trapped on the brane. In our case of one thick brane
immersed in a bulk with infinite extra dimension, if we allow measurements
only at $z=0$, the evidence of Dirac fermions is lost. In the more realistic
scenario, one must consider the finite size of the brane and a nonzero
overlap between the parity odd wavefunctions with the SM fields \cite{gross}%
, which would be sensitive to to the extra dimension in the vicinity of the
brane, justifying the procedure followed in this section.

\subsection{Dimensional analysis and phenomenology}

After restoring the dimensional parameters in the action(\ref{action}%
), it reads as
\begin{equation}
S=\int dx^{4}dy\sqrt{\left\vert g\right\vert }\left[ -\frac{1}{4G^{\left(
5\right) }}R+\frac{1}{2}g^{ab}\partial _{a}\phi \partial _{b}\phi +\frac{1}{2%
}g^{ab}\partial _{a}\chi \partial _{b}\chi -V\left( \phi ,\chi \right) %
\right] .
\end{equation}%
In a five-dimensional space-time, the mass dimension for the fields
and coupling constants in the action are: $\left[ \phi \right] =\left[ \chi %
\right] =M^{3/2}$, $\left[ V\left( \phi ,\chi \right) \right] =M^{5}$, $%
\left[ g_{ab}\right] =1$, $\left[ R\right] =M^{2}$ and the five-dimensional
gravitational constant has dimension $\left[ G^{\left( 5\right) }\right]
=M^{-3}$.

The potential of the scalar fields (\ref{gpot}) is written as
\begin{equation}
V\left( \phi ,\chi \right) =\frac{1}{8}\left( \frac{\partial W}{\partial
\phi }\right) ^{2}+\frac{1}{8}\left( \frac{\partial W}{\partial \chi }%
\right) ^{2}-\frac{1}{3}G^{\left( 5\right) }W^{2}
\end{equation}
where $W$ is the superpotential with mass dimension 4. The coupling $%
G^{\left( 5\right)}$ in the last term guarantees the possibility of the
model to have some compatibility with the expected LHC phenomenology.

In that way, the Bloch brane is defined by the following
superpotential
\begin{equation}
W=2\kappa Z\phi -\frac{2\kappa }{3Z}\phi ^{3}-\frac{2a\kappa }{Z}\phi \chi
^{2}
\end{equation}%
with $\left[ Z\right] =M^{3/2},~\left[ \kappa \right] =M$ and $\left[
a\right] =1$. It yields the following solutions for the scalar fields
\begin{equation}
\phi =Z\tanh \left( 2a\kappa y\right) ~\ \ ,~\ \ \ \ \chi =Z\left( \frac{1}{a%
}-2\right) ^{1/2}\cosh ^{-1}\left( 2a\kappa y\right) ,  \label{kink}
\end{equation}%
where $Z$, the kink amplitude, is given in terms of the
five-dimensional gravitational constant as $Z^{-2}=G^{\left( 5\right) }$
and, the parameter $\kappa ^{-1}$ is related with the brane width. Thus, we
have only two free parameters in the model $\kappa $ and $G^{\left( 5\right)
}$.

From the explicit form of the potential in term of the scalars fields
\begin{eqnarray}
V\left( \phi ,\chi \right) &=& \frac{\kappa ^{2}}{2Z^{2}}\left( Z^{2}-\phi
^{2}-a\chi ^{2}\right) ^{2}+ \frac{2a^{2}\kappa ^{2}}{Z^{2}}\phi ^{2}\chi
^{2}-\frac{4\kappa ^{2}G^{\left( 5\right) }}{3Z^{2}}\phi ^{2}\left( Z^{2}-%
\frac{1}{3}\phi ^{2}-a\chi ^{2}\right) ^{2},
\end{eqnarray}%
we get the masses for the fields $\phi$ and $\chi$ as being
\begin{equation}
m_{\phi }=\sqrt{\frac{14}{3}}\kappa~\ \ ,~\ \ \ m_{\chi }=\sqrt{2a}\kappa,
\end{equation}
therefore, the parameter $\kappa$ is explicitly related with the
field masses. It offers an opportunity for $\kappa$ to be the energy scale
of the model.

The action for massless Dirac fermion is given by
\begin{equation}
S_{f}=\int dx^{4}dy\sqrt{\left\vert g\right\vert }\left[ \frac{{}}{{}}\bar{%
\Psi}\Gamma ^{a}D_{a}\Psi -\gamma \eta \bar{\Psi}F\left( \phi ,\chi \right)
\Psi \right] ,
\end{equation}%
where $\gamma $ is a dimensionless parameter and $\left[ \Psi \right] =M^{2}$%
. As the function $F\left( \phi ,\chi \right) =\phi \chi $ has dimension 3,
we have $\left[ \eta \right] =M^{-2}$.

In order to obtain an expression for the effective coupling constant
$\eta $ in terms of our two free parameters, we check the linear equations
of motion (\ref{alpha1}) and (\ref{alpha2}) after the change of variables $%
\alpha _{L}=e^{-2A}L$ and $\alpha _{R}=e^{-2A}R$:
\begin{eqnarray}
\left[ \partial _{z}+\gamma \eta e^{A}F\left( \phi ,\chi \right) \right] L
&=&mR,  \label{Lap} \\[-0.3cm]
&&  \notag \\
\left[ \partial _{z}-\gamma \eta e^{A}F\left( \phi ,\chi \right) \right] R
&=&-mL.  \label{Rap}
\end{eqnarray}%
By performing the following rescaling:
\begin{equation}
z=\lambda \xi ~,~\ \ \ \partial _{z}=\lambda ^{-1}\partial _{\xi },
\end{equation}%
with $\left[ \lambda \right] =M^{-1}$, the Eqs. (\ref{Lap}%
)-(\ref{Rap}) are rewritten as
\begin{eqnarray}
\left[ \partial _{\xi }+\gamma e^{A}{F}\left( \bar{\phi},\bar{\chi}\right) %
\right] L &=&\bar{m}R \\[-0.3cm]
&&  \notag \\
\left[ \partial _{\xi }-\gamma e^{A}{F}\left( \bar{\phi},\bar{\chi}\right) %
\right] R &=&-\bar{m}L,
\end{eqnarray}%
where $\bar{\phi}=\phi /Z$, $\bar{\chi}=\chi /Z$
are dimensionless fields. We have set 
\begin{equation}
\lambda ^{-1}=\eta Z^{2},
\end{equation}%
and $\bar{m}=m\lambda $ is a dimensionless constant, this
way, the parameter $\lambda ^{-1}$ gives our mass or energy scale.

The choosing of the coupling constant $\eta $ is a fine-tuning
procedure so that the model has the chance of providing a phenomenology in
agreement with LHC energy scale. Thus, setting $\eta =\kappa G^{\left(
5\right) }$, the mass scale results
\begin{equation}
\lambda ^{-1}=\kappa,
\end{equation}%
therefore the mass scale is related to the brane width and
it is independent of the relation between $G^{\left( 5\right)
}$ and the four-dimensional $G$ constant. 

From the kink equations (\ref{kink}), we define the effective brane
width as
\begin{equation}
\Delta L=\frac{1}{4a\kappa }~\ .
\end{equation}%
In order for resonances to appear, our model imposes the condition $%
a\lesssim 0.17$, this gives $\Delta L\gtrsim 1.47\kappa ^{-1}$%
. By considering $\kappa $$\sim 1$ TeV, we
obtain that the brane width is $\gtrsim 10^{-18}$ cm. It is easy
to note that if $\kappa >1$ TeV we get better lower limit for the
brane width.

\section{Conclusions}

In this work we considered branes constructed with two scalar fields. We
considered a simple Yukawa coupling between the two scalars and the spinor
field. Depending on the amount of the coupling parameter $a$ between the
scalars, the energy momentum density of the branes can be characterized by a
two-peak or by a single peak. After \cite{bg} we consider this two-peak
distribution as characterizing a brane with internal structure. For these
more complex branes, the coupling parameter $a$ is small. We investigated
the occurrence of massive modes with both chiralities solving numerically
the Schr\"odinger equation and looking for zero-modes and possible
resonances. For branes with internal structure, we found left- and
right-handed resonances together with a zero-mode left-handed solutions. The
absence of zero-mode solutions for right-handed fermions together with their
findings for left-handed ones agrees with the well-known fact that massless
fermions must be single-handed in a brane model \cite{rpu}. With respect to
resonances for left-handed fermions, we can resume our conclusions with the
following points: i) Larger values of $a$ correspond to larger peaks of
resonance, with corresponding smaller lifetimes. ii) Above a threshold value
of the parameter $a$, the resonances become too unstable and cease to
appear. This coincides with the qualitative change of the energy momentum
density of the brane. iii) Branes with internal structure are more effective
in trapping left-fermions. iv) The lower is the parameter $a$, the higher is
the mass of the the resonant mode. This means that branes with internal
structure tend to trap matter with larger mass more efficiently in
comparison to branes without internal structure. A further investigation on
this subject could consider other classes of Yukawa interaction in order to
look for possible effects on the massive fermionic resonances.

\section{Acknowledgements}

The authors thank L. Losano, D. Bazeia and F. A. Brito for discussions and CNPq,
CNPq/MCT/CT-Infra and PADCT/MCT/CNPq for financial support. The
authors also thank Yu-Xiao Liu and W. Bietenholtz for comments and the
referees for important remarks that improved the first version of this work.

\end{document}